\documentclass[useAMS,usenatbib]{mn2e}
\usepackage{psfig}  
\usepackage{xcolor}
\usepackage{color}

\usepackage{amstext,comment}
\usepackage{amssymb}
\usepackage{amsmath}
\usepackage{float}
\usepackage{appendix}
\usepackage{graphicx}
\usepackage{txfonts}
\usepackage{amsmath,mathtools,revsymb,amssymb}
\usepackage{booktabs,multirow,calc,xspace,rotating}
\usepackage{soul}

\soulregister\cite7
\soulregister\citet7
\soulregister\citep7
\soulregister\ref7
\soulregister\pageref7

\usepackage{scalerel}
\usepackage{tikz}
\usetikzlibrary{svg.path}

\definecolor{orcidlogocol}{HTML}{A6CE39}
\tikzset{
  orcidlogo/.pic={
    \fill[orcidlogocol] svg{M256,128c0,70.7-57.3,128-128,128C57.3,256,0,198.7,0,128C0,57.3,57.3,0,128,0C198.7,0,256,57.3,256,128z};
    \fill[white] svg{M86.3,186.2H70.9V79.1h15.4v48.4V186.2z}
                 svg{M108.9,79.1h41.6c39.6,0,57,28.3,57,53.6c0,27.5-21.5,53.6-56.8,53.6h-41.8V79.1z M124.3,172.4h24.5c34.9,0,42.9-26.5,42.9-39.7c0-21.5-13.7-39.7-43.7-39.7h-23.7V172.4z}
                 svg{M88.7,56.8c0,5.5-4.5,10.1-10.1,10.1c-5.6,0-10.1-4.6-10.1-10.1c0-5.6,4.5-10.1,10.1-10.1C84.2,46.7,88.7,51.3,88.7,56.8z};
  }
}

\newcommand\orcidicon[1]{\href{https://orcid.org/#1}{\mbox{\scalerel*{
\begin{tikzpicture}[yscale=-1,transform shape]
\pic{orcidlogo};
\end{tikzpicture}
}{|}}}}

\DeclareRobustCommand{\ion}[2]{%
\relax\ifmmode
\ifx\testbx\f@series
{\mathbf{#1\,\mathsc{#2}}}\else
{\mathrm{#1\,\mathsc{#2}}}\fi
\else\textup{#1\,{\mdseries\textsc{#2}}}%
\fi}

\usepackage{hyperref}

\newcommand{\begit}{\begin{itemize}}
\newcommand{\enit}{\end{itemize}}
\newcommand{\begen}{\begin{enumerate}}
\newcommand{\enen}{\end{enumerate}}

\newcommand{\beq}{\begin{equation}}
\newcommand{\eeq}{\end{equation}}
\newcommand{\beqa}{\begin{eqnarray}} 
\newcommand{\eeqa}{\end{eqnarray}}

\begin{document}
\label{firstpage}
\title[Mass-Loaded Hot Supersonic Outflows]{Mass-Loading and Non-Spherical Divergence in Hot Galactic Winds: Implications for X-ray Observations}
\author[Dustin D. Nguyen \& Todd A. Thompson]{Dustin D.~Nguyen $^{1,3}$\thanks{E-mail: nguyen.1971@osu.edu} \orcidicon{0000-0002-1875-6522} \& 
Todd A.~Thompson $^{2,3}$\thanks{E-mail: thompson.1847@osu.edu} \orcidicon{0000-0003-2377-9574}\\
  $^{1}$Department of Physics, Ohio State University, 191 W. Woodruff Ave, Columbus, OH 43210 \\
  $^{2}$Department of Astronomy, Ohio State University, 140 W.~18th Ave, Columbus, OH 43210 \\
  $^{3}$Center for Cosmology and Astro-Particle Physics, Ohio State University, 140 W.~18th Ave, Columbus, OH 43210 \\}

\date{Accepted XXX. Received YYY; in original form ZZZ}
\maketitle

\begin{abstract}
Cool clouds are expected to be destroyed and incorporated into hot supernova-driven galactic winds. The mass-loading of a wind by the cool medium modifies the bulk velocity, temperature, density, entropy, and abundance profiles of the hot phase relative to an un-mass-loaded outflow. We provide general equations and limits for this physics that can be used to infer the rate of cool gas entrainment from X-ray observations, accounting for non-spherical expansion. In general, mass-loading flattens the density and temperature profiles, decreases the velocity and increases the entropy if the Mach number is above a critical value. We first apply this model to a recent high-resolution galactic outflow simulation where the mass-loading can be directly inferred. We show that the temperature, entropy, and composition profiles are well-matched, providing evidence that this physics sets the bulk hot gas profiles. We then model the diffuse X-ray emission from the local starburst M82. The non-spherical (more cylindrical) outflow geometry is directly taken from the observed X-ray surface brightness profile. These models imply a total mass-loading rate that is about equal to that injected in the starburst, $\simeq 10$\,M$_\odot$ yr$^{-1}$, and they predict an asymptotic hot wind velocity of $\sim 1000\,{\rm km \ s^{-1}}$ that is $\sim1.5-2$ times smaller than previous predictions. We also show how the observed entropy profile can be used to constrain the outflow velocity, making predictions for future missions like XRISM. We argue that the observed X-ray limb-brightening may be explained by  mass-loading at the outflow's edges. 
\end{abstract}

\section{Introduction}
\label{sec:intro}
Understanding the nature of galactic outflows is fundamental to understanding how galaxies evolve. Winds are commonly seen in star-forming galaxies at both low and high redshifts \citep{Martin1998,Pettini2001,Rubin2010,Heckman2015,Heckman2016}. Galactic winds modulate star formation, shape both the stellar mass function and the mass-metallicity relation \citep{Dekel1986,Finlator2008,Peeples2011}, and advect metals beyond the host galaxy into both the circumgalactic medium (CGM) and intergalactic medium (IGM) \citep{Aguirre2001,Scannapieco2002,Tremonti2004,Oppenheimer2006,Oppenheimer2008}.  

Galactic winds are multi-phase and multi-dynamical. X-ray observations of nearby starforming galaxies such as M82 reveal a hot phase at $T \approx 10^6 - 10^7 $ K \citep[e.g,][]{Strickland2007}. Optical and UV observations reveal ionized and neutral atomic $\approx 10^4-10^5$ K gas \citep{Heckman1990,Martin2005,Westmoquette2009}. Far-infrared, sub-mm, and molecular tracers reveal the cold $\leq 10^4$ K medium \citep{Walter2002,Leroy2015}. These different temperature components are observed to have different kinematics. In some systems the cold phase is observed to be gravitationally bound and is recycled as a fountain-like flow \citep{Leroy2015}, whereas in others the the velocity exceeds the escape velocity and is able to leave the host galaxy.

\citet{Chevalier1985} provided the analytic solution for a wind driven by uniform energy and mass injection that undergoes adiabatic expansion as it leaves the driving region (hereafter CC85). The model requires input of two physical observables (out of $v, \ n, \, \  \mathrm{and} \ T$) in order parameterize the dimensionless quantities $\alpha$ and $\beta$, which describe the amount of energy and mass-injection within the central region. The hot phase kinematics have not been directly observed yet. Previous X-ray observations of M82's central temperature and densities constrained the energy and mass-injection to values of $0.3 \leq \alpha \leq 1$ and $1 \leq \beta \leq 2.8$, respectively, which predict that the hot supersonic gas is moving at $v  = 1400-2000 \ \mathrm{km \ s^{-1}}$ \citep{Strickland2009}. 

A critical open issue in galactic wind physics is the acceleration mechanism for cool and warm gas in outflows and their interactions with the hot phase. The prevailing picture is that winds are driven by shock-heated hot gas produced by stellar winds and supernovae (SNe), as predicted by the CC85 model. However, it has been shown that clouds embedded in a hot supersonic medium undergo hydrodynamical shredding before being accelerated by ram pressure to observed velocities \citep{Cooper2009,Scannapieco2015,Zhang2017,Schneider2018}. \citet{Gronke2018} showed there exists a parameter space where cold clouds will actually grow and become accelerated to high velocities due to a cooling and mixing cycle. However, the growth of cold clouds has yet not been observed in global SNe-driven galactic wind simulations \citep{Schneider2020}. 

Mixing of cold, dense, material in a hot supersonic flow is expected to affect its thermodynamics and kinematics. CGOLS IV found that energy and momentum mixing between the hot and cold phases leads to radial profiles of the hot gas that do not undergo evolution indicative of adiabatic spherical expansion \citep{Schneider2020}. New observations of M82, which capture the diffuse X-rays 2.5 kpc above and below the plane, too, have temperature and density profiles that fall flatter than adiabatic spherical expansion. The presence of mixing in CGOLS IV and M82, as well its essential role in cloud growth \citep{Gronke2018}, all motivate a study on mixing of a cool phase into a hot outflow in the context of galactic winds its implications for observations. 

Additional physics have been included to SNe-driven winds such as radiative cooling \citep{Wang1995,Thompson2016,Lochhaas2020}, the inclusion of cosmic rays and radiation pressure \cite{Yu2020}, and both additional mass-loading and wind-collimation outside the driving region \citep{Suchkov1996}. Our study deviates from \cite{Suchkov1996} by explicitly including non-spherical divergence into the analytics of our 1D wind model (whereas they deduced results from 2D simulations) and a simpler description of mass-loading (we do not include a physical description of the cold clouds as \citet{Cowie1981,Suchkov1996,Fielding2021} do). This allowed us to derive equations that explicitly show how both geometry and mass-loading affect the evolution of the hot outflow (see Equations \ref{eq:dvds2}, \ref{eq:drhods2}, and \ref{eq:dTds2}). Furthermore, we present a prescription on interpreting X-ray observations, where we posit that the outflow geometry should be directly gathered from analysis tools, such as XSPEC for Chandra data, which can be used in our flexible models to then infer the mixing rate. This is important because it is not self-consistent to use a tool like XSPEC to determine the central temperature and density, which may take on a non-spherical specific flow geometry during analysis, and then to use a spherical CC85 wind model or any other arbitrary flow geometry to make predictions about the hot gas kinematics and thermodynamics. We apply our models to CGOL IV and M82 data and show mass-loading and/or non-spherical divergence may be able to explain the deviations from a spherical adiabatic wind as observed by \citet{Schneider2020} and \citet{Lopez2020}. 

The paper proceeds as follows: In \autoref{sec:two} we briefly review the CC85 wind model and present the hydrodynamic equations that embody this work. We derive equations that directly show how observables are affected by both the mixing of a hot and cool material as well as the affects due to geometry of the hot outflow itself. In \autoref{sec:three} we investigate the effects of mass-loading and non-spherical divergence independently in two case-studies that better reveal the implications of the derived equations. Relative to an adiabatic spherical flow, both mass-loading and non-spherical divergence lead to flatter temperature, pressure, and density profiles. Mass-loading decelerates the hot outflow as kinetic energy is thermalized and provides heating. Adiabatic supersonic winds undergoing spherical expansion and flow through flared-cylinder geometry do not decelerate. However, other rapidly varying geometries can decelerate the flow, and we derive the criterion for these flow geometries. The entropy profile typically increases in a mass-loaded wind whereas it remains constant for a non-radiative flow, regardless of flow geometry. However, we show that there is a critical Mach number below which the entropy of the flow decreases if mass is added. We use this criterion to derive the minimum/maximum gas velocity for increasing/decreasing entropy. Lastly, mass-loading leads to a decreasing abundance profile, as the hot metal-rich flow is mixed with metal poor cool gas. In \autoref{sec:four} we apply our models to both CGOLS IV data and to recent observations of M82. For the CGOLS IV comparison, the mass-loading of a supersonic flow increases temperatures, densities, and pressure, while decreasing the velocity and the "scalar" (abundance-like quantity), in a way that is qualitatively similar to the simulation results. For the M82 comparison, we find that a mass-loaded flow into the non-spherical flow geometry, directly determined by XSPEC, provide an explanation for the deviations of a spherical adiabatic wind in the observed temperatures, densities, X-ray luminosity, and metallicity profiles. 

\section{General Wind Equations} 
\label{sec:two} 
\subsection{The CC85 Model} 
In the CC85 model, mass and energy from SNe are uniformly injected in a sphere of radius $R$ at rates:
\begin{align}
    \dot{M}_\mathrm{hot} & = \beta \ \dot{M}_\mathrm{SFR}, \label{eq:mdothot} \\ 
    \dot{E}_\mathrm{hot} & = 3.1 \times 10^{41} \ \mathrm{ergs \ s ^{-1}} \ \alpha \ (\dot{M}_\mathrm{SFR} / \mathrm{M_\odot \ yr^{-1}}),
    \label{eq:edothot}
\end{align}
where $\dot{M}_\mathrm{SFR}$ is the star formation rate, $\alpha$ and $\beta$ are the energy thermalization and mass-loading efficiencies, respectively, and that it is assumed that there is one core-collapse SN per 100 $\mathrm{M}_\odot$ of star formation and that each SN seeds the galaxy with with $10^{51}$ ergs. The edge of the driving region $r=R$ has the characteristic solution that $\mathcal{M}=1$. The CC85 model provides an analytic solution to this model problem assuming spherical adiabatic expansion outside $R$. For specified $\alpha$, $\beta$, $\dot{M}_\mathrm{SFR}$, and $R$, the classical CC85 solution provides velocity, density, and temperature solutions for the full range of radii under the assumption that the affects of radiative cooling and gravity are negligible. 

For the purpose of this paper, we use the classic CC85 solution for $r<R$, but then add additional physics such as optically-thin radiative cooling, mass-loading from a cool phase, and non-spherical divergence for $r>R$ along the wind axis. We do not explicitly solve for solutions at the sonic point $r=R$, but instead focus on the modifications in the supersonic wind region. 

\subsection{Radiative Cooling} 
\label{sec:cooling}
If a hot wind is sufficiently mass-loaded within the starburst region (i.e, high values of $\beta$), then the radiative cooling time for the outflow can be shorter than the dynamic timescale so that the hot wind undergoes rapid cooling into a high velocity cold gas \citep{Wang1995,Silich2003,Silich2004,Thompson2016}. To capture this physics we include cooling for regions outside the driving nucleus ($r>R$). However we note that cooling in the interior region may be important, as for sufficiently high values of $\beta$ the flow may become radiative within the core which lowers the overall mass outflow rate \citep{Wunsch2007,Wunsch2008,Lochhaas2020}. In the calculations presented, we adopt the cooling curve given by \citet{Schneider2018}, which is an analytic fit to a solar-metallicity collisional ionization equilibrium curve calculated with the Cloudy simulation by \citet{Ferland2013}. This curve assumes a temperature floor of $10^4$ K to account for heating and photoionization from UV sources. Throughout the paper we take $\mu = 0.6$ for a plasma with solar metallicity.

\subsection{Hydrodynamic Equations}

\begin{figure}
    \centering
    \includegraphics[width=\columnwidth]{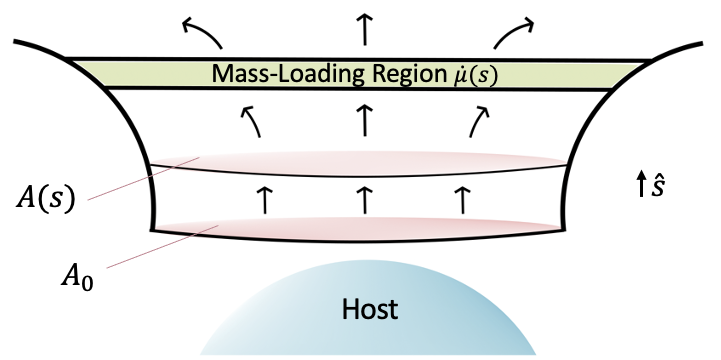}
    \caption{The geometry of the flared-cylinder flow tube for the 1D wind model. As detailed in the CC85, the flow undergoes spherical expansion within the host region. The area function then takes the form $A(s)$ at $r \geq R$.}
    \label{fig:geo}
\end{figure}

We wish to investigate the question of what happens when a hot, energy-driven, supersonic galactic flow sweeps up matter in the form of cold clouds. The physical picture is that cold clouds from the ISM are entrained into the hot flow outside of the wind driving region at $r>R$. These clouds are crushed and shredded by hydrodynamical instabilities, being incorporated into the hot gas over a time and length scale set by the cloud crushing time \citep[e.g][]{Klein1994,Zhang2017}. For a flow moving in the $\hat{s}$ direction, the hydrodynamic equations are 

\begin{align}
    & \frac{1}{A} \frac{d}{ds} ( \rho v A ) = \dot{\mu}, \label{eq:continuity} \\ 
    &  v \frac{dv}{ds} = - \frac{1}{\rho} \frac{dP}{ds} - \nabla \Phi - \frac{\dot{\mu} v}{\rho}, \label{eq:momentum} \\
    &  v \frac{d\epsilon}{ds} - \frac{ v P}{\rho^2 } \frac{d\rho}{ds} = - \frac{n^2 \Lambda}{\rho} + \frac{\dot{\mu}}{\rho} \bigg( \frac{v^2}{2} - \epsilon - \frac{P}{\rho} \bigg), \label{eq:energy} 
\end{align}
where $v$, $\rho$, $P$, $n$, $\epsilon$, $\Lambda$, $\nabla \Phi$ are the velocity, mass density, pressure, number density, specific internal energy, cooling function, and gravitational acceleration along $\hat{s}$, respectively. Similar to that derived by \citet{Cowie1981}, these equations assume that cold material is injected with zero temperature and zero velocity with respect to the hot phase - which is valid in the limit that the hot phase contains most of the kinetic and thermal energy. The term in parentheses in the internal energy equation represents the thermalization of kinetic energy in sweeping up the cold medium. Writing Equation \ref{eq:continuity} in differential form as
\begin{equation}
    \frac{d \ln \rho }{ds} + \frac{d \ln v}{ds} + \frac{ d \ln A}{ds} = \frac{\dot{\mu}}{\rho v}, 
\end{equation}
we find that 
\begin{align}
    \frac{1}{v} \frac{dv}{ds} (v^2 - c_s^2) = - \frac{\dot{\mu}}{v \rho} (v^2 + & c_s^2) + c_s^2 \frac{d\ln A}{ds} - \frac{d \Phi}{ds} \nonumber \\ 
    & \frac{1-\gamma}{v} \bigg[ - \frac{n^2 \Lambda}{\rho} + \frac{\dot{\mu}}{\rho} \bigg( \frac{v^2}{2} - \epsilon - \frac{P}{\rho} \bigg) \bigg], 
\end{align}
\begin{align}
    \frac{1}{\rho} \frac{d\rho}{ds} (v^2 - c_s^2) =  \frac{2 v \dot{\mu}}{\rho}   -  v^2  & \frac{d\ln A}{ds} + \frac{d\Phi}{ds} \nonumber \\ & + \frac{(\gamma -1 )}{v} \bigg[ - \frac{n^2 \Lambda}{\rho} + \frac{\dot{\mu}}{\rho} \bigg( \frac{v^2}{2} - \epsilon - \frac{P}{\rho} \bigg) \bigg],
\end{align}
and 
\begin{align}
    \frac{dT}{ds} (v^2 - c_s^2 ) = \frac{c_T^2}{C_V} \bigg( & \frac{2v \dot{\mu}}{\rho} -  v^2 \frac{d \ln A}{ds} + \frac{ d\Phi}{ds} \bigg) \nonumber \\ 
    & + \frac{(v^2 - c_T^2)}{v C_V} \bigg[ - \frac{n^2 \Lambda}{\rho} + \frac{\dot{\mu}}{\rho} \bigg( \frac{v^2}{2} - \epsilon - \frac{P}{\rho} \bigg) \bigg],
\end{align}
where $C_V$, $c_s$, and $c_T$ are heat capacity, adiabatic sound speed, and isothermal sound speed respectively. These so-called ``wind equations" show how a single component flow in a pipe of arbitrary shape reacts to mass addition, optically thin radiative heating/cooling, and a given gravitational potential along the individual stream line. From these equations, one could derive the shift in the sonic point associated with these effects by setting the right hand side to zero in each case. If gravity and radiative cooling can be neglected, the equations can be rewritten in terms of the Mach number as: 
\begin{align}
    & \frac{d \ln v}{ds} (\mathcal{M}^2 -1 ) = - \frac{\gamma \dot{\mu} \mathcal{M}^2}{v \rho} + \frac{d \ln A}{ds} , \label{eq:dvds1} \\ 
    & \frac{d \ln \rho}{ds} (\mathcal{M}^2 -1) = \frac{ [(1+\gamma)\mathcal{M}^2 -1 ] \dot{\mu}}{v \rho} - \mathcal{M}^2 \frac{d\ln A}{ds}, \label{eq:drhods1} \\ 
    & \frac{d \ln T}{ds} (\mathcal{M}^2 -1) =  \frac{\dot{\mu}}{2 v \rho}[\gamma (\gamma-1)\mathcal{M}^4 +  (\gamma -3) \mathcal{M}^2 +2] \nonumber \\
    & \hspace{5cm} - (\gamma-1) \mathcal{M}^2 \frac{d \ln A}{ds}. \label{eq:dTds1}
\end{align}
The entropy gradient can be written as 
\begin{equation}
    \frac{1}{k_b} \frac{d \ln K}{ds} = \frac{d \ln T }{ds} + (1-\gamma) \frac{d \ln n}{ds} 
\end{equation}
which for a non-radiative flow, with negligible gravity, becomes
\begin{align}
    \frac{d \ln K}{ds} = \frac{k_b \dot{\mu}}{v \rho} \frac{[\gamma(\gamma-1)\mathcal{M}^4 + (\gamma - 2 \gamma^2 - 1) \mathcal{M}^2  + 2 \gamma]}{2 (\mathcal{M}^2 -1 )}.
    \label{eq:dlnKdsGammas}
\end{align}
Note, unlike the wind-equations for $v$, $\rho$, and $T$ (Equations \ref{eq:dvds1}, \ref{eq:drhods1}, and \ref{eq:dTds1}) that there is no explicit dependence on $A(s)$ (as we would expect). For $\gamma=5/3$ this becomes
\begin{equation}
    \frac{d \ln K }{ds} =  \frac{k_b \dot{\mu}}{v \rho} \frac{ \bigg( \frac{5}{9} \mathcal{M}^4 - \frac{22}{9} \mathcal{M}^2 + \frac{5}{3} \bigg)}{(\mathcal{M}^2 -1 )}. 
    \label{eq:dlnKds}
\end{equation}
In the highly supersonic limit, $\mathcal{M} \gg 1$, and for $\gamma = 5/3$ and smoothly varying flows the equations above simplify to: 
\begin{align}
    & \frac{d \ln v}{ds} = - \frac{5}{3} \frac{\dot{\mu}}{v \rho},  \label{eq:dvds2} \\ 
    & \frac{d \ln \rho}{ds} = \frac{8}{3} \frac{\dot{\mu}}{v \rho} - \frac{d \ln A}{ds}, \label{eq:drhods2} \\ 
    & \frac{d \ln T }{ds} =  \frac{5}{9} \mathcal{M}^2 \frac{\dot{\mu}}{v \rho} - \frac{2}{3} \frac{d \ln A}{ds}. \label{eq:dTds2} \\ 
    & \frac{d \ln K}{ds} = \frac{5}{9} \frac{k_b \dot{\mu}}{v\rho} \mathcal{M}^2 \label{eq:dKds2}
\end{align}
From these equations it is apparent that mass-loading leads to flatter temperature and density profiles, decelerates the hot gas, and increases its entropy. For a spherical flow $s \rightarrow r$, $A \propto r^2$, and $d \ln A / dr = 2 / r $.  
By setting $\dot{\mu} \rightarrow 0$, the expressions accord with expectation for supersonic adiabatic spherical flow (i.e., $v = \ \mathrm{const}$, $\rho \propto r^{-2}$, $T \propto r^{-4/3}$, and $K = \mathrm{const}$).
\subsection{Implications of Non-Zero Entropy Gradients}
\label{sec:entropygrad}
We see from Equations \ref{eq:dKds2}, that mass-loading increases the entropy of a hot supersonic flow. However, there are instances where the entropy will instead decrease. Solving for the roots in Equation \ref{eq:dlnKds}, we derive the critical Mach number\footnote{For $\gamma=4/3$, $\mathcal{M}_\mathrm{crit} \simeq 2.5$} at which the gradient of the entropy will either be positive or negative due to mass-loading:
\begin{align}
    \mathcal{M}_\mathrm{K,crit} &= \sqrt{\frac{1}{5}(11+\sqrt{46})} \\  
   & = 1.89 \hspace{1.5cm} (\mathcal{M} > 0.92).
    \label{eq:Mcrit}
\end{align}
For a supersonic flow and $\mathcal{M} < \mathcal{M}_\mathrm{K,crit}$ the entropy profile gradient is negative due to mass-loading, and will be positive for $\mathcal{M} > \mathcal{M}_\mathrm{K,crit}$. Measurements of the temperature, can be used to derive the sound speed of the gas $c_s$ which can then be substituted into Equation \ref{eq:Mcrit}. Observations of a negative or positive entropy gradient can then be used to establish a maximum or minimum gas velocity. Once $T$ is measured,  
\begin{equation}
    v_\mathrm{max,\,min}(r)  = {\mathcal M}_{\rm K,\,crit} \sqrt{\frac{5 k_bT(r)}{3 \mu m_p}}
    \simeq 900\,{\rm km\,\,s^{-1}}\,\,\left(\frac{T}{10^7\,{\rm K}}\frac{0.6}{\mu}\right)^{1/2}
    \label{eq:v}
\end{equation}
where the constraint provides a maximum on the velocity if the entropy gradient is negative and a minimum on the velocity if the entropy gradient is positive, assuming that the Mach number is larger than unity.\footnote{We note that there is another root to equation (\ref{eq:dlnKds}), $\mathcal{M}_\mathrm{K,crit} = 0.92$. However in our work presented here, we focus on the supersonic flow outside of the wind-driving region.} We discuss this limit in our application to numerical simulations and observations of the starburst M82 in Section \ref{sec:four}.

\section{Results} 
\label{sec:three}

As encapsulated in Equations \ref{eq:dvds2}, \ref{eq:drhods2}, \ref{eq:dTds2}, and \ref{eq:dKds2} both mass-loading and the flow geometry affect the kinematic and thermodynamic evolution of the outflow. We  first investigate the effects of these two components independently and we neglect gravity for simplicity.

\subsection{Non-Spherical Adiabatic Expansion}
\label{sec:nonsphericaladiabatic}

\begin{figure}
    \centering
    \includegraphics[width=\columnwidth]{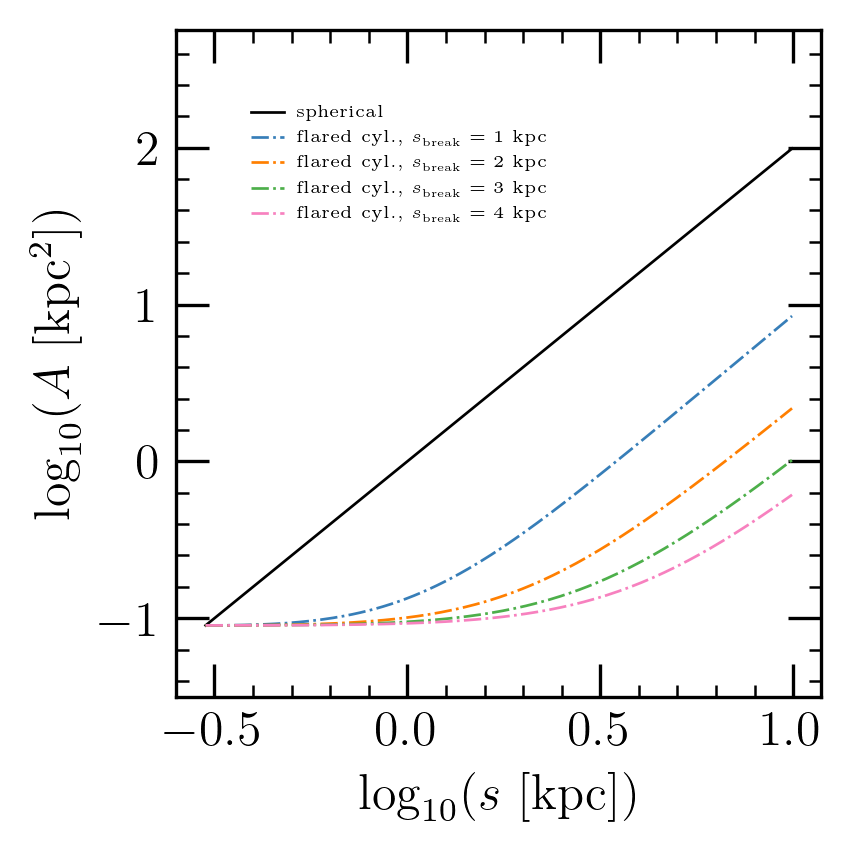}
    \caption{The flared-cylinder area function (colored dotted-dashed lines) for constant cylindrical area and varied pipe heights (see Equation \ref{eq:flaredcylarea}). The black line is spherical expansion. After the cylindrical flows pass $s_\mathrm{break}$, they undergo spherical expansion and thus acquire the same log-log slope of 2.}
    \label{fig:flaredareas}
\end{figure}

\begin{figure*}
    \centering
    \includegraphics[width=\textwidth]{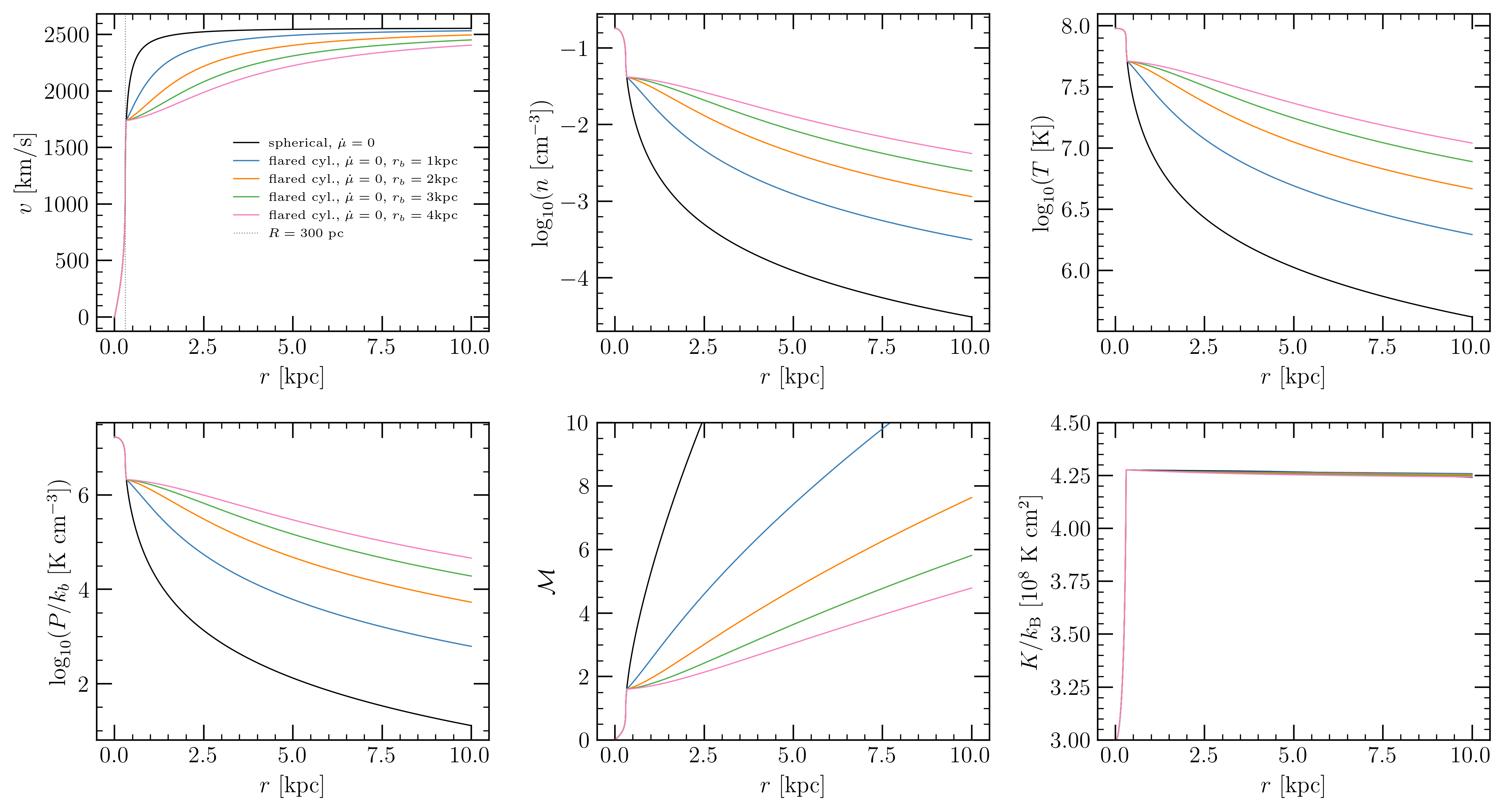}
    \caption{Velocity, density, temperature, pressure, Mach number, and entropy for a representative set of solutions for a wind expanding into a flared-cylinder geometry for varied pipe heights $s_\mathrm{break}$ (see eq.~\ref{eq:flaredcylarea}). All models have have the same CC85 parameters: $\alpha = 1$, $\beta = 0.15$, $\dot{M}_\mathrm{SFR}=10 \ \mathrm{M}_\odot \ yr^{-1}$, $R=0.3$\,kpc; and cylindrical base area $A_0 = R^2$ with a volumetric mass-loading rate of $\dot{\mu} = 0$.}
    \label{fig:flaredcyl}
\end{figure*}

We first consider the flared-cylinder geometry which has been used in previous studies on coronal flux tubes and two-component wind studies of the Milky Way \citep[e.g,][]{Breitschwerdt1991,Everett2008}. We use the CC85 model to provide the initial conditions of the problem, focusing on the supersonic regime. We assume the flow undergoes uniform energy and mass injection and spherical divergence in the region $r < R$, as illustrated by Figure \ref{fig:geo}, and then begins non-spherical expansion at $r \geq R$ described by $A(s)$.  The area function for a flared cylinder is 
\begin{equation}
    A(s) = A_0 \ [1+ (s/s_\mathrm{break})^2],
    \label{eq:flaredcylarea}
\end{equation}
where $s_\mathrm{break}$ is the approximate height at which the flow transitions from a cylinder with area $A_0$ to that of a spherical wind. In Figure \ref{fig:flaredareas} we plot the areas for varied pipe heights $s_\mathrm{break}=1, \ 2, \ 3, \ \mathrm{and} \ 4$ kpc, and fixed base area $A_0 = R^2$ with $R=0.3$\,kpc.
Figure \ref{fig:flaredcyl} shows velocity, density, temperature, pressure, Mach number, and entropy solutions for varied pipe heights of $s_\mathrm{break} = 1, \ 2, \ \mathrm{and} \ 3$\,kpc. Motivated by previous studies of starburst M82 \citep[e.g,][]{Strickland2009,Lopez2020}, all fiducial models assume M82-like CC85 parameters: $\alpha = 1$, $\beta = 0.15$, $\dot{M}_\mathrm{SFR}=10 \ \mathrm{M_\odot \ yr^{-1}}$, and $R = 0.3$\,kpc. 

Relative to the adiabatic spherical expansion model, we see that a flow that undergoes non-spherical expansion, in this case through a flared-cylinder, has shallower temperature, density, and pressure profiles as suggested by Equations \ref{eq:dvds2}, \ref{eq:drhods2}, and \ref{eq:dTds2}. As expected, the entropy initially increases due to energy and mass injection within the starburst region (CC85, $r < R$) but remains approximately constant as it undergoes adiabatic expansion ($r \geq R$) regardless of the outflow geometry (see Equation \ref{eq:dlnKds}).

For illustration, we contrast the flared cylinder geometry (Equation \ref{eq:flaredcylarea}) with a fluted cylinder geometry which was first used in studies of coronal holes \citep{Kopp1976}. The area function takes the form 
\begin{equation}
    A(s) = s^2 \ f(s) 
    \label{eq:flutedcylarea}
\end{equation}
where the function $f(s)$ is the non-radial expansion factor: 
\begin{equation}
    f(s) = \frac{ f_\mathrm{max} e^{(s-S_1)/\sigma} +f_1}{e^{(s-S_1)/\sigma} +1},
    \label{eq:flutedfs}
\end{equation}
with
\begin{equation}
    f_1 = 1 - ( f_\mathrm{max} -1 ) e^{(S_0 - S_1)/\sigma}.
    \label{eq:flutedf1}
\end{equation}
The function $f(s)$ varies most rapidly with distance at $s = S_1$ , with most of the change occurring in the narrow region between $S_1 - \sigma $ and $S_1 + \sigma$. $S_0$ is the starting distance at which non-spherical divergence occurs. $f_\mathrm{max} = 1$ corresponds to a spherical geometry ($s^2$); and $f_\mathrm{max} > 1$ and $f_\mathrm{max} < 1$ correspond to slower and faster than $s^2$ divergence within the expansion region, respectively.  

\begin{figure*}
    \centering
    \includegraphics[width=\textwidth]{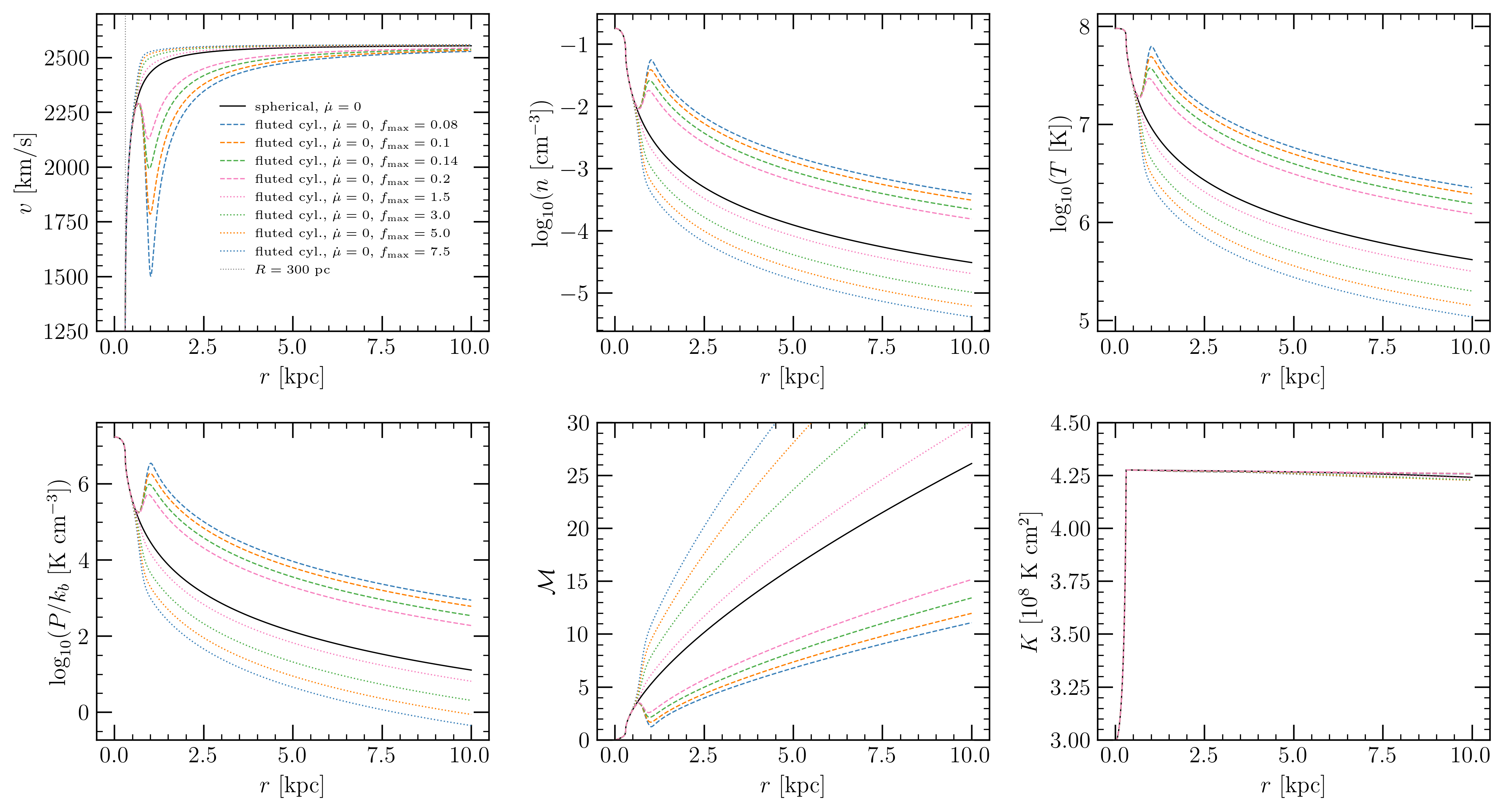}
    \caption{Velocity, density, temperature, pressure, Mach number, and entropy solutions to the general equations for varied geometries $f_\mathrm{max} = 0.08, \ 0.1, \ 0.14, \ 0.2, \ 1.5, \ 3.0, \ 5.0, \ \mathrm{and} \ 7.5$ (see eqs.~\ref{eq:flutedcylarea}, \ref{eq:flutedfs}, and \ref{eq:flutedfs}). All models assume M82-like CC85 parameters: $\alpha = 1$, $\beta = 0.15$, $\dot{M}_\mathrm{SFR}=10 \ \mathrm{M}_\odot \ yr^{-1}$, and $R = 0.3$\,kpc; and $\sigma =0.2 R$, $S_1 = 2.5 R$, $S_0 = R$}
    \label{fig:flutedcyl}
\end{figure*}

Figure \ref{fig:flutedcyl} shows velocity, density, temperature, pressure, Mach number, and entropy solutions for varied geometries $f_\mathrm{max} = 0.08, \ 0.1, \ 0.14, \ 0.2, \ 1.5, \ 3.0, \ 5.0, \ \mathrm{and} \ 7.5$. Again, all fiducial models assume M82-like CC85 parameters: $\alpha = 1$, $\beta = 0.15$, $\dot{M}_\mathrm{SFR}=10 \ \mathrm{M}_\odot \ yr^{-1}$, and $R = 0.3$\,kpc; and geometry parameters: $\sigma =0.2 R$, $S_1 = 2.5 R$, and $S_0 = R$. 

Relative to the adiabatic spherical expansion model, we see that a flow that undergoes expansion in a fluted cylinder has shallower temperature, density, and pressure profiles for $f_\mathrm{max} > 1$ and steeper profiles for $f_\mathrm{max} < 1$. However, unlike the flared-cylinder geometry, we see deceleration and acceleration of the gas despite a constant entropy profile. This highlights the importance of the $d \ln A / ds$ term in Equation \ref{eq:dvds1}, which was negligible (Equation \ref{eq:dvds2}) in the flared cylinder case shown in Figure \ref{fig:flaredcyl}. By setting the velocity gradient to zero in Equation \ref{eq:dvds1}, the criterion for non-negligible acceleration (and deceleration) of a supersonic flow due to rapidly changing flow geometry is derived \ref{eq:dvds1} to be  
\begin{equation}
    \bigg | \frac{d \ln A}{d \ln s} \bigg | / (\mathcal{M}^2 -1 ) \gg 1.  \hspace{2.5cm} (\mathcal{M} > 1) 
    \label{eq:decelGeo}
\end{equation}

\subsection{Mass-Loading of a Spherical Wind} 
\label{sec:massloadedspherical}

\begin{figure*}
    \centering
    \includegraphics[width=\textwidth]{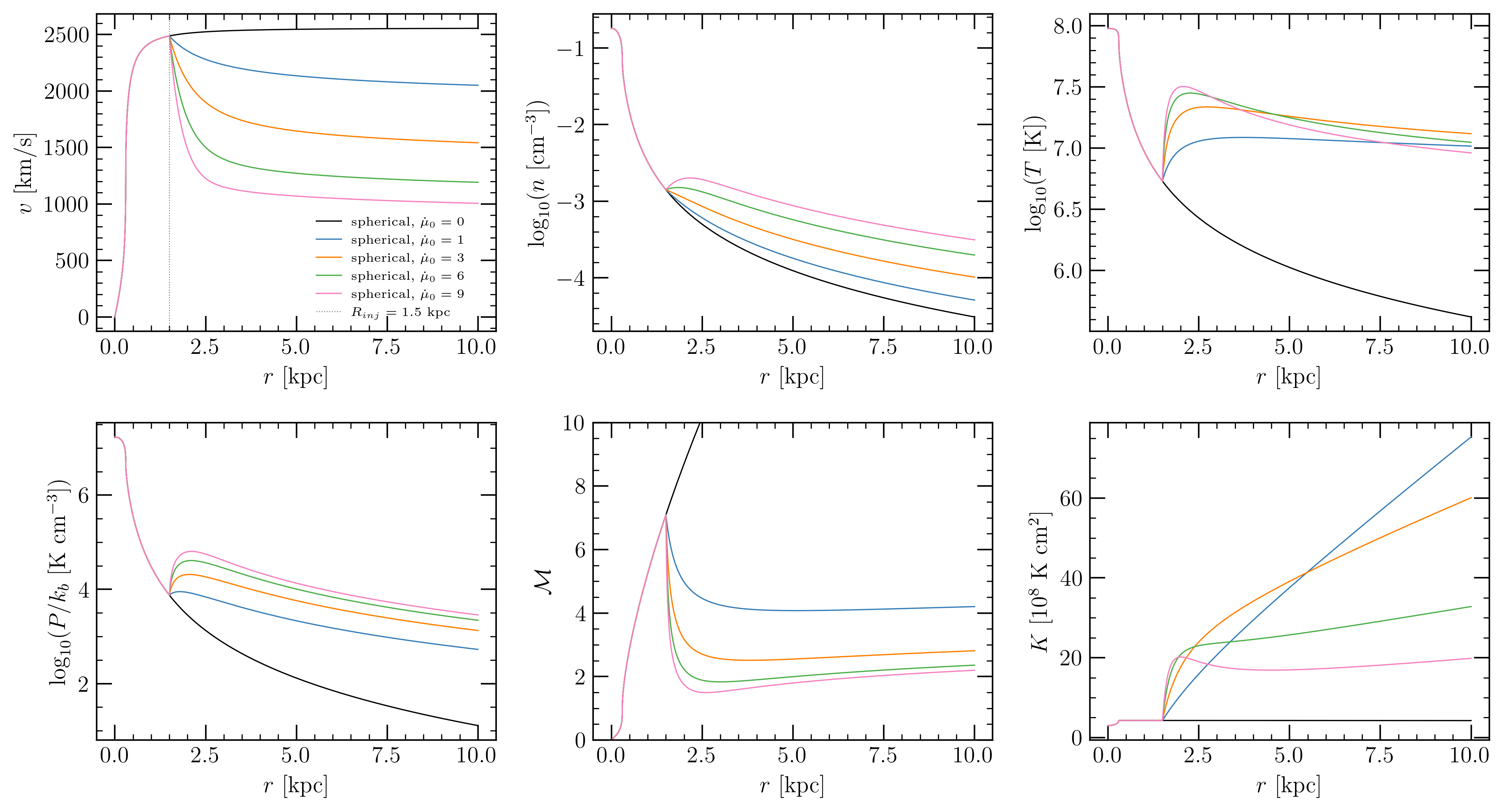}
    \caption{Velocity, density, temperature, pressure, Mach number, and entropy for a representative set of solutions for a spherical wind with varied volumetric mass-loading rates $\dot{\mu}_0= 0, \ 1, \ 3, \ 6, \ \mathrm{and} \ 9 \ \mathrm{M_\odot \ kpc^{-3} \ yr^{-1}}$ (see eq.~\ref{eq:mudot}). All models have have the same CC85 parameters: $\alpha = 1$, $\beta = 0.15$, $\dot{M}_\mathrm{SFR}=10 \ \mathrm{M}_\odot \ yr^{-1}$, $R=0.3$\,kpc, and $R_\mathrm{load} = 1.5$\,kpc; and spherical area function $A(r) \propto r^2$ and mass-loading power-law index of $\Delta = 3.1$.}
    \label{fig:mudot}
\end{figure*}

We now explore the effects of mass loading occurs in a spherical supersonic wind. We assume that begins to be deposited outside the starburst nucleus at radius $R_\mathrm{load}$ and that the volumetric mass-loading rate takes the form
\begin{equation}
    \dot{\mu} = \dot{\mu}_0 \  (R_\mathrm{load}/r)^\Delta,  \quad \quad (r \ge R_\mathrm{load})  
    \label{eq:mudot}
\end{equation}
where $\dot{\mu}_0$ has units of $\mathrm{M_\odot \ kpc^{-3} \ yr^{-1}}$. The total mass loading rate is
\begin{equation}
    \dot{M}_\mathrm{load} = 4\pi \int_{R_\mathrm{load}}^r dr \ \dot{\mu} \ r^2.
\end{equation}
For $\dot{M}_\mathrm{load}$ to be comparable to $\dot{M}_\mathrm{hot}$, we then roughly require that
\begin{align}
    \dot{\mu}_0 &= \frac{3}{4\pi} \frac{\beta \dot{M}_\mathrm{SFR}}{R_\mathrm{load}^3} \\ 
                &\simeq 5 \ \mathrm{M_\odot \ kpc^{-3} \ yr^{-1}} \ \bigg( \frac{\beta \dot{M}_\mathrm{SFR}}{20 \,\mathrm{M_\odot / yr}} \bigg) \bigg( \frac{ \mathrm{kpc} }{R_\mathrm{load}} \bigg)^3. 
\end{align}
We define a quantity 
\begin{equation}
     c(r) = \frac{\dot{M}_\mathrm{hot}}{\dot{M}_\mathrm{hot} + \dot{M}_\mathrm{load}}, 
     \label{eq:scalar}
\end{equation}
which tracks how much mass has been added to the flow. In this form $c(r) \rightarrow 1$ if $\dot{\mu} =0$. If $c(r) = 0.5$, then half of that mass at $r$ is due to mass loading.

Figure \ref{fig:mudot} shows velocity, density, temperature, pressure, Mach number, and entropy solutions for $\dot{\mu}_0= 0, \ 1, \ 3, \ 6, \ \mathrm{and} \ 9 \ \mathrm{M_\odot \ kpc^{-3} \ yr^{-1}}$. All fiducial models have the same CC85 parameters as in Figures \ref{fig:flaredcyl} and \ref{fig:flutedcyl} using the flared and fluted cylinder geometries: $\alpha = 1$, $\beta =0.15$, $\dot{M}_\mathrm{SFR}=10 \ \mathrm{M}_\odot \ yr^{-1}$, $R = 0.3$\,kpc, and $R_\mathrm{load}=1.5$\,kpc. All models in Figure \ref{fig:mudot} assume spherical geometry with $A = r^2$, and the power-law index for Equation \ref{eq:mudot} is $\Delta = 3.1$.

Relative to the adiabatic spherical expansion model, we see that a spherical flow that undergoes mass-loading has shallower temperature, density, and pressure profiles as suggested by Equations \ref{eq:dvds2}, \ref{eq:drhods2}, and \ref{eq:dTds2}. The injected mass decelerates the flow, thermalizing its kinetic energy and heating the flow. The entropy initially increases due to energy and mass injection within the starburst region (CC85) and continues to increase as more mass is seeded into the hot outflow (Equation \ref{eq:dKds2}). The exception is the heavily mass-loaded model (pink line), where the entropy initially increases, decreases, and then increases again. This is due to the critical Mach number condition, Equation \ref{eq:Mcrit}. The flow initially has $\mathcal{M}>\mathcal{M}_\mathrm{crit}$ at $R_{\rm load}$, but $\dot{\mu}_0$ is large enough that it is strongly decelerated by mass loading until $\mathcal{M}<\mathcal{M}_\mathrm{crit}$ at $\simeq2$\,kpc. The entropy then decreases until $\mathcal{M}>\mathcal{M}_\mathrm{crit}$ at $\simeq4$\,kpc and the entropy increases thereafter.

Comparing Figures \ref{fig:flaredcyl} and \ref{fig:flutedcyl} with Figure \ref{fig:mudot}, we note that while both a cylindrical geometry and mass loading can produce much flatter temperature and density profiles than in the adiabatic spherical case, the entropy gradient is a key discriminator. Mass-loading leads directly to a  non-zero entropy gradient, whereas changes in outflow geometry maintain constant entropy. Thus, $dK/ds$ is a critical observational diagnostic of mass loading.

\subsection{Radiative Winds from Mass-Loading}
\label{sec:Radwinds}
\begin{figure*}
    \centering
    \includegraphics[width=\textwidth]{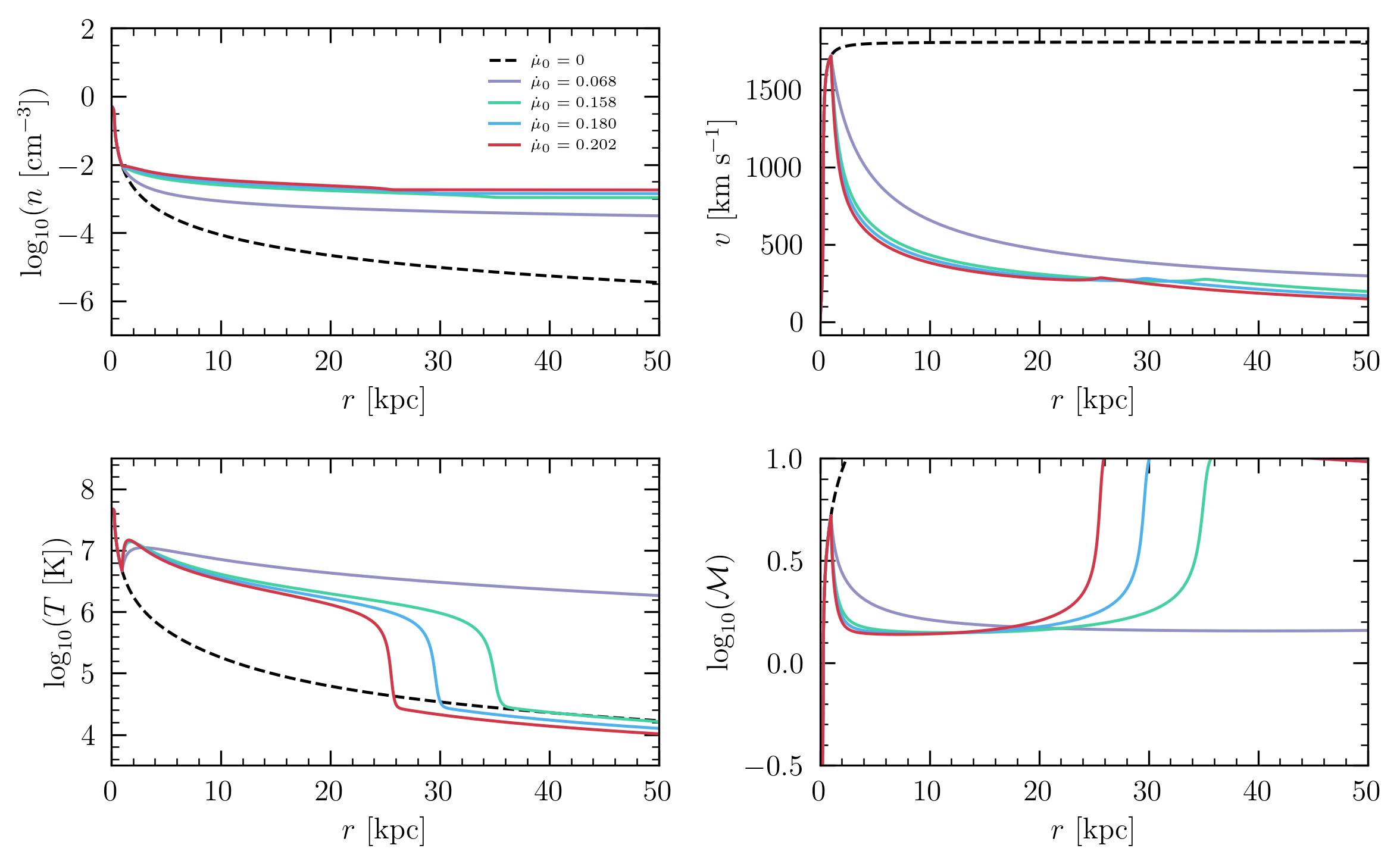}
    \caption{Density, velocity, temperature, and Mach number profiles for a spherical wind with varied mass-loading rates: $\dot{\mu}_0 = 0, \ 0.069, \ 0.158, \ 0.180,\ 0.202 \ \mathrm{M_\odot \ kpc^{-3} \ yr^{-1}}$ (see eq.~\ref{eq:mudot}). All models have the same CC85 parameters: $\alpha = 1$, $\beta = 0.3$, $R=0.3$\,kpc, $\dot{M}_\mathrm{SFR} = 10 \ \mathrm{M_\odot \ yr^{-1}}$, $\Delta=2$, and $R_\mathrm{load}=1.0$\,kpc. A CC85 wind with $\alpha=1$ and $\beta=0.3$ is non-radiative (black dashed line). However, we see that with sufficient mass-loading (red, green, and blue lines), the initially non-radiative wind can undergo rapid cooling.} 
    \label{fig:radwinds}
\end{figure*}

Previous studies have considered radiative winds that arise from heavy mass-loading within the wind-driving region of a galactic wind model \citep{Wang1995,Silich2003,Thompson2016}. In particular, \cite{Thompson2016} show that for $\beta\gtrsim1$, that the medium becomes radiative on kpc scales. However, rather than heavy mass-loading within the host galaxy (i.e., high $\beta$), an initially non-radiative wind (i.e., low $\beta$) can still become radiative if it is sufficiently mass-loaded outside of the wind driving region. In Figure \ref{fig:radwinds}, we consider an initially non-radiative wind with a range of volumetric mass-loading rates of $\dot{\mu} = 0.0$, 0.068, 0.158, 0.180, 0.202\,$\mathrm{M_\odot \ kpc^{-3} \ yr^{-3}}$, and other wind parameters of $\alpha = 1$, $\beta = 0.3$, R = 0.3\,kpc, $\dot{M}_\mathrm{SFR} = 10 \ \mathrm{M_\odot \ yr^{-1}}$, $\Delta=2$, and $R_\mathrm{load} = 1.0$\,kpc (see eq.~\ref{eq:mudot}).

The models with high mass-loading (red, blue, and green lines) exhibit rapid cooling while models with low mass-loading (purple) and no mass-loading (black dashed) do not. In order to explore even stronger mass loading and small cooling radii, we require a numerical treatment of the problem that can pass through the sonic point.

Before turning to applications of these models to the CGOLS IV simulation and observations of M82, we note that the details of the profiles shown in Figure \ref{fig:mudot} are strongly affected by $R_{\rm load}$, $\dot{\mu}_0$, and $\Delta$. The models shown here are meant to be illustrative of the behavior when the mass-loading is peaked at small radii ($\Delta>3$) and with low enough $\beta$ that radiative cooling is not strong (compare with \citealt{Thompson2016}).

\section{Application}
\label{sec:four}
\subsection{CGOLS IV} 
\label{sec:cgols}
CGOLS IV is an extremely high resolution, 3D, simulation employing a unique cluster feedback scheme to drive energetic hot winds from a high surface density galactic disk, which gave rise to a complex multiphase outflow \citep{Schneider2020}. The simulation was carried out in a box with uniform grid cells of dimension $10 \times 10 \times 20$\,kpc with constant cell width of 4.9\,pc.

Among their results, \cite{Schneider2020} find that the density, pressure, and temperature profiles of the hot phase along the wind axis do not match the expectation for a purely adiabatic spherical wind. The black dashed lines in Figure \ref{fig:cgols} show the profiles from Figure 6 of \cite{Schneider2020} at a snapshot in time (35\,Myr). The gray solid line shows the expectation for a spherical adiabatic flow. The density, temperature, and pressure exhibit much flatter profiles than expected, while the velocity shows a strong deceleration near the base of the outflow and a slow rise to a smaller asymptotic velocity on kpc scales. Simultaneously, the entropy increases, and the ``scalar" $c$, which measures the mixing between the hot wind fluid ($c=1$) and the cool swept up material ($c=0$), strongly decreases. 

Section \ref{sec:massloadedspherical} and Figure \ref{fig:mudot} all suggest that these thermodynamic and kinematic deviations away from a simple adiabatic wind can be explained by mass-loading from the cool phase into the hot phase.  Indeed, \citet{Schneider2020} are able to explain the dynamics of the cool gas using mixing of momentum and energy from the hot phase. Here, we use the expressions from Section \ref{sec:massloadedspherical} to understand the effect of mass-loading on the bulk properties of the hot phase. 

\begin{figure}
    \centering
    \includegraphics[width=\columnwidth]{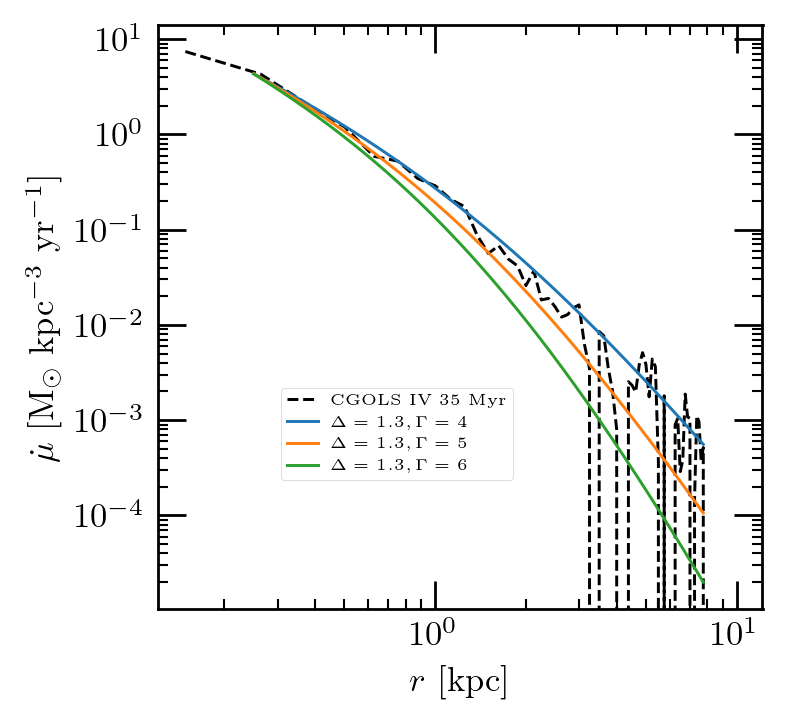}
    \caption{The volumetric mass-loading rate $\dot{\mu}$ in the hot phase as inferred from CGOLS IV data (black dashed line). We plot three different smoothened-two power law models (blue, orange, and green solid lines).}
    \label{fig:infermudot}
\end{figure}

\begin{figure*}
    \centering
    \includegraphics[width=\textwidth]{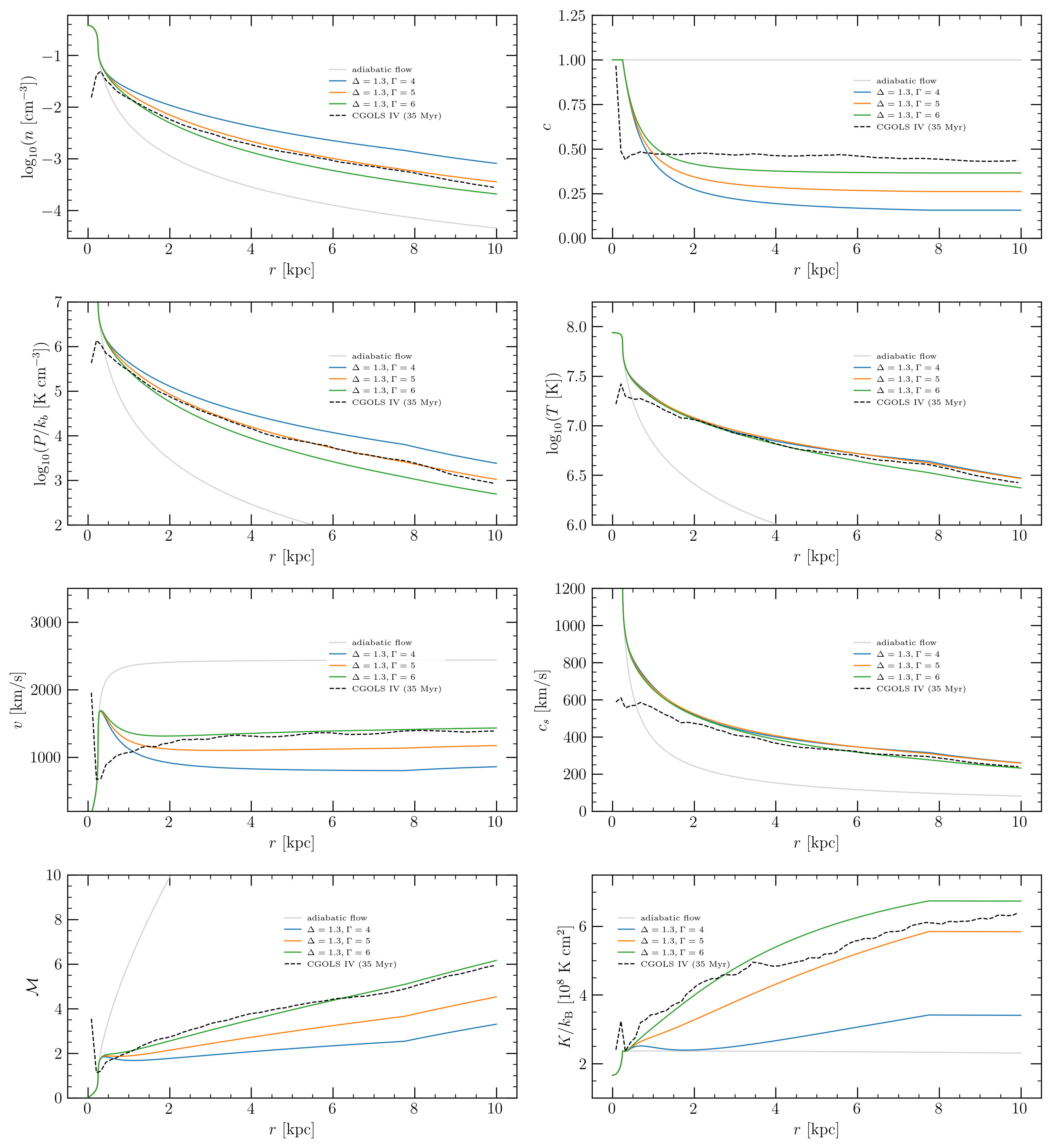}
    \caption{Density, mass-tracking variable $c(r)$, pressure, temperature, velocity, adiabatic sound speed, Mach number, and entropy solutions for four models: 1 spherical adibatic wind model and 3 mass-loaded spherical wind models with varied mass-loading rates as described by Figure \ref{fig:infermudot}. The CGOLS IV data is denoted by the black dashed line and the colored solid lines are the models. }
    \label{fig:cgols}
\end{figure*}

In focusing on only the hot phase, we first directly infer the volumetric mass-loading rate $\dot{\mu}$ as a function of radius using the velocity and density profiles from the equation of continuity (Eq.~\ref{eq:continuity}). We then fit two-component power laws of the form 
\begin{equation}
    \dot{\mu} =  \dot{\mu}_0 \ \frac{a^\Delta}{r^\Delta ( 1+ r /a)^{\Gamma-\Delta}}  \quad \quad (r \ge R_\mathrm{load}), 
    \label{eq:mudot2power}
\end{equation}
which describes a smooth transition on scale $a$ from $\dot{\mu}\propto r^{-\Delta}$ ($r\ll a$) to $\dot{\mu}\propto r^{-\Gamma}$ ($r\gg a$).  Figure \ref{fig:infermudot} shows the inferred $\dot{\mu}$ from the CGOLS IV simulation (black dashed) and three models with $\Gamma = 4, \ 5, \ 6$ (solid colored), which are all normalized to the same value at $R_\mathrm{load}=R$ such that $\dot{\mu}_0 = 0.64, \ 0.75, \ \mathrm{and} \ 0.87 \ \mathrm{M_\odot \ kpc^{-3} \ yr^{-1} }$, respectively. All models have $\Delta = 1.3$ and $a = 1.5$ kpc.

Given these derived profiles for $\dot{\mu}(r)$, we solve Equations \ref{eq:continuity}, \ref{eq:momentum}, and \ref{eq:energy}, and then compare our models to the results of \citet{Schneider2020}. For simplicity, we neglect non-spherical expansion for this comparison because the wind profiles from \cite{Schneider2020} are computed in a bi-cone of constant opening angle and because the flow appears roughly spherical by-eye on $\gtrsim\,{\rm kpc}$ scales (see their Fig.~4). Note that on scales near the driving region ($\lesssim0.5$\,kpc), the flow is very complicated, with individual cluster outflows interacting with each other and with the surrounding disk material, and for these reasons we expect our one-dimensional model to break down.

As in our examples shown in Figure \ref{fig:mudot}, we use a simple CC85 model for the interior driving region, and thus we do not attempt to match the more complex cluster feedback scheme employed in CGOLS IV. All models shown have the following parameters: $\alpha = 0.85$, $\beta = 0.14$, $\dot{M}_\mathrm{SFR} = 15 \ \mathrm{M_\odot \ yr^{-1}}$, and $R = 0.25$ kpc. Using the inferred volumetric mass-loading rate $\dot{\mu}(r)$ from Figure \ref{fig:infermudot} for the blue, orange and green lines, Figure \ref{fig:cgols} shows the derived density, mass-tracking variable $c(r)$ (Equation \ref{eq:scalar}), pressure, temperature, velocity, adiabatic sound speed, Mach number, and entropy of our models (blue, orange, and green). Gravity is neglected because of the relatively low-mass potential used in the CGOLS IV simulations. The central mass-loading used for these models ($\beta=0.14$) leads to a non-radiative wind, as the critical mass deposition is not met \citep{Thompson2016}. With the additional mass-loading at $R_\mathrm{load}$, the wind is still non-radiative and does not exhibit the rapid cooling discussed in Section \ref{sec:Radwinds}. 

By construction, the lines bracket the density profile of the CGOLS IV simulation on $\sim1-10$\,kpc scales. Remarkably, the temperature, pressure, and entropy profiles are also well-reproduced, with a much flatter temperature profile than would be obtained for standard spherical adiabatic expansion (gray line). As we have emphasized in Section \ref{sec:massloadedspherical}, the positive entropy gradient is a hallmark of mass-loading when the Mach number of the hot flow is greater than $\sim1.9$ (eq.~\ref{eq:dlnKds}). If this was an actual starbust wind with observed entropy and temperature profiles, but with unconstrained hot gas velocity profile, we note that equation (\ref{eq:v}) would immediately imply a minimum for the hot gas velocity since the entropy gradient is positive throughout the region. Using the temperature or sound speed profiles, equation (\ref{eq:v}) implies a decreasing minimum velocity with $v_{\rm min}\simeq1000$\,km s$^{-1}$ at $\simeq2$\,kpc and  $v_{\rm min}\simeq500$\,km s$^{-1}$ at $r\simeq10$\,kpc.

The velocity and Mach number profiles are less well-reproduced by the models. We are unable to match the sudden deceleration of CGOLS IV at very small radii just outside the driving region where the Mach number is close to unity. This is perhaps to be expected, given the complexity of the driving region (multiple star clusters in arranged in a disk-like morphology, in different stages of evolution). While all three mass-loaded models produce nearly flat velocity profiles from $\sim2-10$\,kpc, the $\Delta=5$ (orange) model, which fits the density profile best, under-predicts the velocity across the computed range. Finally, all three models under-predict the $c(r)$ profile on large scales, predicting somewhat more mass-loading than the CGOLS IV simulation suggests. 

We view this as excellent agreement, given the simplicity of the model. In particular, we neglect a number of physical and qualitative effects that surely complicate a direct comparison with the simulations. In particular, our model assumes that the cool, entrained gas has zero bulk velocity, maximizing thermalization of the hot wind as the cooler material is entrained. Similarly, we neglect any coupling of the cool gas with the hot gas so that there is no backreaction of the entrainment process on cool gas (e.g., via acceleration of the cool material). In reality, the simulation shows that the cool material is accelerated throughout the volume and that the dynamics of both phases are coupled. This physics may help explain the large scale differences between the model and the simulation. At small radii, the model does a worse job of reproducing the behavior of the simulation, but this is expected as the simulation has a complex cluster feedback scheme that is inherently non-uniform and not time-steady. The medium further exhibits non-spherical expansion on small scales where the wind coalesces into a single bulk flow, and potentially on larger scales (see Fig.~4 from \citealt{Schneider2020}). A more complete exploration would include dynamical equations for the cool material, acceleration of the cool material and the necessary modifications to the hot wind energy equation. Finally, an analysis of the gas streamlines on kpc scales would reveal if non-spherical expansion quantitatively affects the entropy and Mach number profiles. Nevertheless, the models shown do an excellent job at reproducing the basic effects seen both qualitatively and quantitatively: increasing entropy, decreasing $c$, and much flatter temperature and pressure profiles.

\subsection{X-Rays from M82} 
\label{sec:m82}

We now attempt a preliminary application of the theory of mass-loaded flows to the starburst M82. \citet{Lopez2020} used Chandra X-Ray Observatory imaging and spectra to map the hot plasma properties of M82's multi-phase wind \citep{Leroy2015}. Previous studies focused on hard X-rays in sub 0.5\,kpc nuclear region \citep[e.g,][]{Strickland2007,Strickland2009}. 
Similar to \citet{Ranalli2008}, \cite{Lopez2020} constructed high-resolution images of the diffuse X-ray emission 2.5 kpc above and below the starburst ridge that are approximately 10 times deeper than \citet{Strickland2009}. Aside from the different X-ray telescopes, \citeauthor{Lopez2020}'s work differs from \citeauthor{Ranalli2008} through the inclusion of an additional temperature component, and non-thermal components such as charge-exchange and a power-law component. They found that the flux-weighted hot gas temperature and density profiles fell off much more slowly than would be expected for spherical adiabatic expansion. Moreover, some of the derived metal abundance profiles (notably Si and S) decrease substantially for the first kpc above and below the plane. These shallow temperature and density profiles and steep drops in abundance are reminiscent of the CGOLS IV results that were the focus of the last section (see Figure \ref{fig:cgols}), suggesting that mixing of a metal-rich hot wind with a metal-poor cooler medium outside of the driving starburst region could explain the observations.

To model the profiles obtained by \cite{Lopez2020}, we first need to set the central wind parameters and the outflow geometry. The central temperature and density of a CC85-like wind are given by \citep{Chevalier1985,Strickland2009}: 
\begin{align}
    T_c & = \frac{0.4 \ \mu m_p \dot{E}_\mathrm{hot}}{k_b \dot{M}_\mathrm{hot}}, \label{eq:Tc} \\ 
    n_c & = \frac{0.296 \ \dot{M}_\mathrm{hot}}{\mu m_p \dot{E}_\mathrm{hot}^{1/2} R^2}, \label{eq:nc}  
\end{align}
where $\mu m_p$ is the average particle mass. Equations (\ref{eq:Tc}) and (\ref{eq:nc}) are used to derive the central values of $\alpha$ and $\beta$ from measurements of the central temperature and density. The difference between this work and that of \cite{Strickland2009} is that while we similarly apply Equations (\ref{eq:Tc}) and (\ref{eq:nc}) to the measurements of the central temperature and density, we also attempt to adjust both the outflow geometry and the volumetric mass-loading to match the extended emission 2.5\,kpc above and below the plane. Our work differs from \citet{Suchkov1996} in that they did not have observations of the extended emission, they focused on the mass-loading in the core driving region of the starburst, and they implemented a physical description of cool clouds in a hot medium. 

\begin{figure*}
    \centering
    \includegraphics[width=\textwidth]{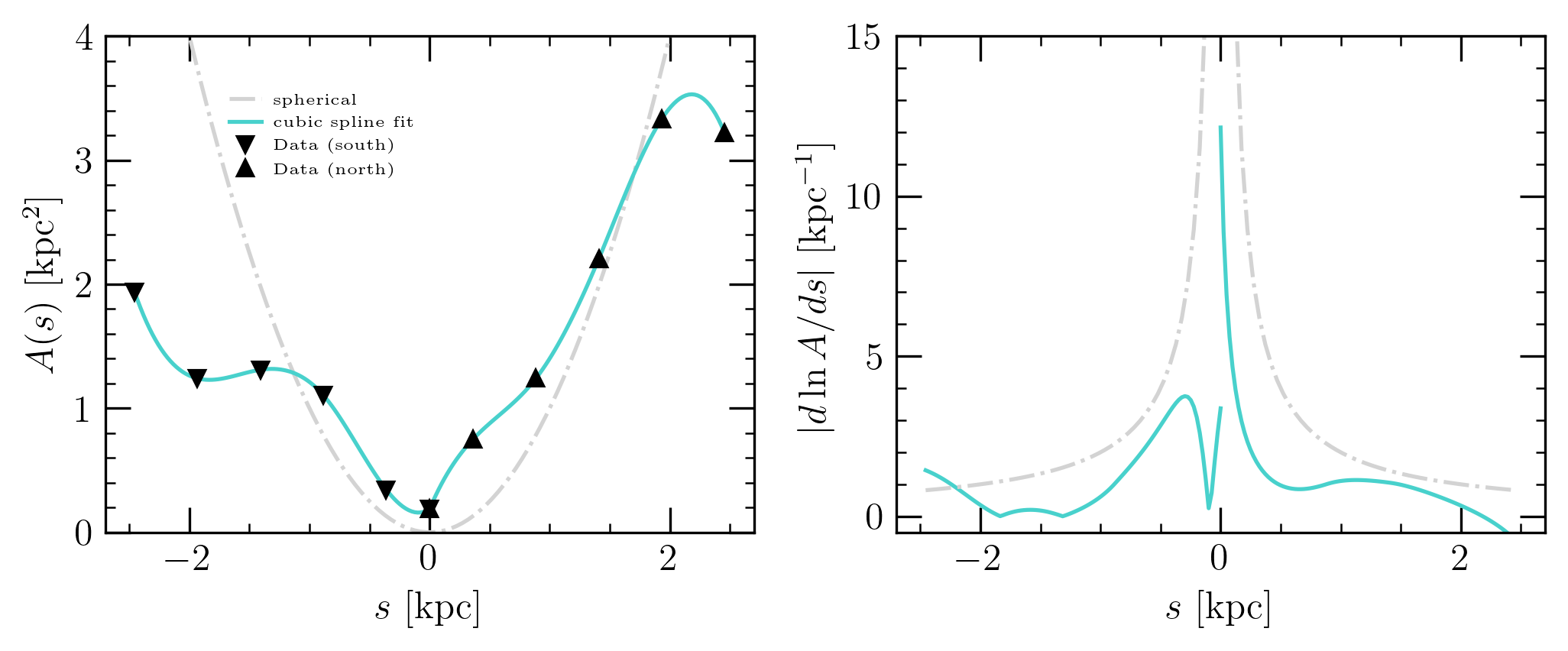}
    \caption{\textit{Left}: The non-spherical area function for M82's hot wind derived from the observed X-ray (0.5-7 keV) surface brightness contours perpendicular to the minor axis as a function of height $s$ above and below the starburst disk at $s=0$ (private communication, L. Lopez).  The black triangles show the lateral size of the contour enclosing 68\% of the emission as a function of distance along the minor axis. The turquoise lines are a cubic spline fit to the southern and northern sides independently. The light gray dotted dashed line is the area element for spherical expansion. \textit{Right}: $d \ln A / ds$ for the cubic spline fits to $A(s)$ and for spherical expansion (gray dot-dashed). This term directly affects the wind profiles (e.g., eqs.~\ref{eq:dvds2}-\ref{eq:dKds2}).}
    \label{fig:M82areas}
\end{figure*}

The outflow geometry is derived directly from the X-ray observations. \citet{Lopez2020} calculate the cylindrical cross section of the hot along the minor axis using the transverse X-ray surface brightness profile as a function of height, with the criterion that 99\% of the emission is enclosed in the cylindrical volume of each region. Figure \ref{fig:M82areas} shows the non-spherical expansion factor $A(s)$ (black triangles) and cubic spline fits to the $A(s)$ measurements shown as the turquoise line in Figure \ref{fig:M82areas}. The $A(s)$ here is determined using 68\% of the enclosed emission (private communication, L. Lopez) rather than the 99\% presented in Table 3 of \citet{Lopez2020}.

\begin{figure*}
    \centering       
    \includegraphics[width=\textwidth]{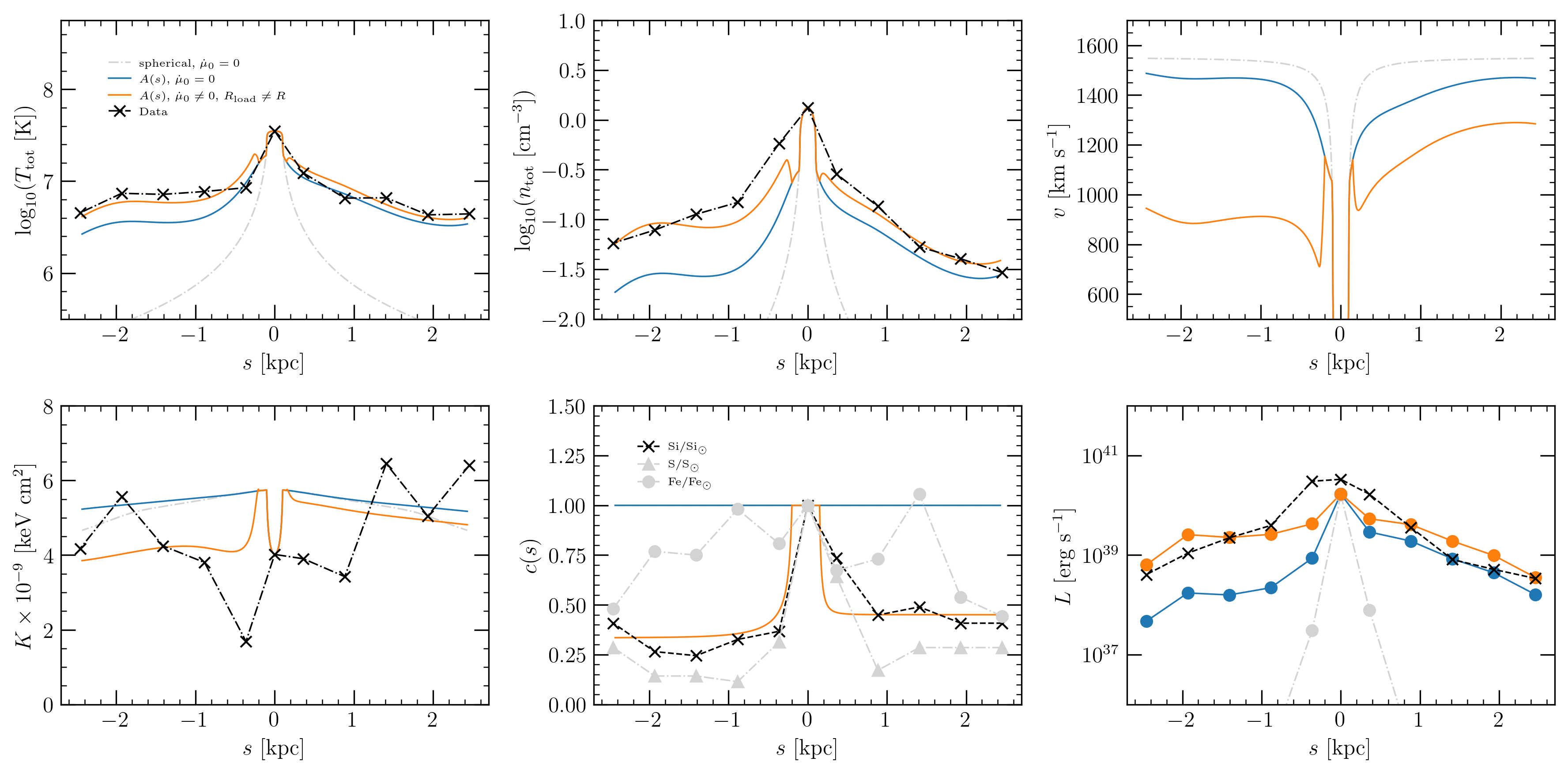}
    \caption{Temperature, density, velocity, entropy, mass-tracking variable $c(r)$, and luminosities for three different models for each wind axis: 1. adiabatic spherical (gray dashed line), 2. adiabatic non-spherical (blue solid), and 3. mass-loaded non-spherical  (orange solid). The non-spherical models have different flow area elements $A(s)$ for each minor axis direction. Similarly, the mass-loaded model has different injection radii and loading rates for each direction. The black dashed line are the density-weighted measurements of M82 observed by Chandra (private communication, L. Lopez).}
    \label{fig:plots}
\end{figure*}

\citet{Lopez2020} used XSPEC to model the X-ray emitting gas as a multi-temperature plasma, finding three individual components with central temperatures of 0.72, 1.46, and 6.89 keV which \citet{Lopez2020} term the ``warm-hot," ``hot," and ``very-hot."  The latter was detected only in the very central region of the starburst. In comparing our models with the data, we are forced to make a choice about how to treat the observed multi-temperature plasma with a single-component mass-loaded hot wind model. For this preliminary comparison with the data, we construct a density-weighted temperature and total density of hot gas as follows:
\begin{equation}
    T_\mathrm{tot} = \frac{\sum_i T_i \cdot n_i}{\sum_i n_i},
    \label{eq:ttot}
\end{equation}
and
\begin{equation}
    n_\mathrm{tot} = \sum_i n_i,
    \label{eq:ntot}
\end{equation}
where $i = 1, \ 2, \ 3$ corresponds to the warm-hot, hot, and very-hot temperature components.  Figure \ref{fig:plots} shows the density, temperature, and entropy inferred within the $A(s)$ 68\% surface brightness contour using these expressions (black dot-dashed lines).

In comparing with the observations, we also calculate the X-ray luminosity of our wind models in the same energy bandwidth of the \citet{Lopez2020} observations ($0.5-7$\,keV) using cooling tables calculated with \texttt{PyAtomDB} \citep{Foster2020} assuming collisional ionization equilibrium (CIE), which is justified for M82 within the central region, where the densities are highest \citep{Strickland2009}. We use this approximation here, but note that the assumption of CIE may be inadequate in less dense regions as the outflow expands and should be assessed with the X-ray emission in future work (in particular, \cite{Gray2019a,Gray2019b} find that the wind medium can be over-ionized). Assuming solar metallicity, the luminosity of a region with volume $A(s)\ ds$ is given by \citep{Zhang2014,Strickland2009}:
\begin{equation}
    L_X^{\nu_1,\nu_2} = \int_{\nu_1}^{\nu_2} \int_{s_1}^{s_2} \ n_e n_H \Lambda(T,\nu) \,A \,\,d\nu\,ds
    \label{eq:lx}
\end{equation}
where $\Lambda$ is emissivity at temperature $T$ and frequency $\nu$ and $n_e$ and $n_H$ are the electron and hydrogen number densities, respectively. The emissivity is calculated over the observing range, 0.5 - 0.7 keV (see Figure \ref{fig:ct}). For comparison with our derived models, the lower right panel of Figure \ref{fig:plots} shows the total X-ray luminosity (including all components) from \cite{Lopez2020} (black dot-dashed line) calculated with an emissivity that is integrated over the observing band.

Solving the general equations (\ref{eq:continuity}), (\ref{eq:momentum}), and (\ref{eq:energy}) using the cubic-spline fit for $A(s)$ determined from Figure \ref{fig:M82areas}, we derive profiles of temperature, density, velocity, entropy, mass-tracking variable $c$ (eq.~\ref{eq:scalar}), and luminosity  along both the southern and northern wind axes. Figure \ref{fig:plots} shows three models, with each adding different pieces of physics. All models assume the same CC85 parameters in the central region: $\alpha=0.18$, $\beta = 0.07$, which are derived from the central temperature and density of the hot component using Equations \ref{eq:Tc} and \ref{eq:nc} for $\dot{M}_\mathrm{SFR} = 10$\,$\mathrm{M_\odot \ yr^{-1}}$ and $R=0.1$ kpc. The gray dot-dashed lines in Figure \ref{fig:plots} shows an adiabatic spherical CC85-like calculation obtained without using the general $A(s)$ or mass-loading. The blue line shows an adiabatic non-spherical with the general $A(s)$ from Figure \ref{fig:M82areas} and with $\dot{\mu}=0$. Note that the blue curve is not symmetric between north and south because of $A(s)$. The orange solid line shows a general non-spherical and mass-loaded model, which includes both $A(s)$ and $\dot{\mu}(s)$ for the northern and southern regions. The functional form of the mass-loading function is given by Equation \ref{eq:mudot2power}. For the southern side, $R_\mathrm{load} = 0.2$ kpc, $\dot{\mu}_0 = 0.36 \ \mathrm{M_\odot \ kpc^{-3} \ yr^{-1}}$, $a = 0.7$ kpc, and $\Delta=\Gamma=3.5$. For the northern side, $R_\mathrm{load}=0.15$ kpc, $\dot{\mu}_0 = 0.01 \ \mathrm{M_\odot \ kpc^{-3} \ yr^{-1}}$, $a=0.7$ kpc, and $\Delta = \Gamma = 5$. 

As discussed in \cite{Lopez2020}, the spherical, non-radiative, wind model (gray) is a poor fit to the data. In contrast, the model with non-spherical $A(s)$ derived from the observations in Figure \ref{fig:M82areas} (blue) does a much better job, especially on the northern side where the density and temperature can be nearly matched with zero mass-loading. Adding mass-loading (orange) makes the temperature and density gradients even flatter, lowers the predicted velocity profile, increases the X-ray luminosity, decrease $c(s)$, and leads to a complicated entropy profile. Overall, this comparison suggests that non-spherical areal divergence and mass-loading may well explain the extended X-ray emission of M82. There are a few important pieces to highlight and several caveats:

{\bf Abundances:} In the middle lower panel of Figure \ref{fig:plots} we plot the Si/Si$_\odot$  (black crosses), S/S$_\odot$ (gray triangles), and Fe/Fe$_\odot$ (gray circles) abundances from \cite{Lopez2020}, normalized to their central values for comparison with $c(s)$, which tracks the relative contribution of the injected wind ($c=1$) and the entrained material ($c=0$). Interestingly, the same $\dot{\mu}$ profile we used to approximate the $T$ and $n$ profiles produces a $c(s)$ profile that roughly matches the decreasing Si abundances as a function of height $s$ above the starburst plane. This could in principle be understood if highly Si-rich central hot wind material is mixed with very Si-poor cool material. Given the definition of $c(s)$, the Si abundance of the swept up material would have to be near zero for the comparison to be apt. S also shows a steep gradient, but with lower relative fractional abundance on kpc scales. In contrast, the Fe abundance profile is more constant. Interpreted within this model, these observations suggest that the Fe abundance of the swept up material is closer to that of the host wind, and that the S and Si abundances of the swept up medium are very small.

{\bf Velocity profile:} We emphasize that the velocity profile is directly affected by the mass-loading, as made clear by equation (\ref{eq:dvds2}): mass-loading in general decreases the velocity.  As shown in the upper right panel of Figure \ref{fig:plots}, the asymptotic velocity obtained from both the spherical and non-spherical wind models (gray and blue) is approximately $1500-1600$\,km s$^{-1}$, whereas the mass-loaded models have terminal velocities of $\sim900$\,km s$^{-1}$ to the south and $\sim1300$\,km s$^{-1}$ to the north. The heavier mass-loading to the south in this model predicts a significantly lower asymptotic velocity than would be predicted by models like \cite{Strickland2009}. That mass-loading may be partially responsible for the flatter temperature and density profiles and the complicated entropy profile seen in the data thus predicts lower hot wind velocities that may be accessible to upcoming missions like XRISM \citep{XRISM2020}. We return to empirical constraints on hot gas velocity in our discussion of the entropy profile below.

{\bf X-ray luminosity:} The lower right panel of Figure \ref{fig:plots} shows the total luminosity profile from \cite{Lopez2020}, which includes all of the thermal components, and components from charge-exchange and non-thermal power-law emission. Using equation (\ref{eq:lx}), the emission from the models is provided for comparison. The non-spherical mass-loaded model (orange) produces more emission than  both the spherical adiabatic and non-spherical adiabatic models. As the flow is mass-loaded, its density increases, which  increases the cooling rate, and thus the luminosity. Relative to a spherical adiabatic model, the non-spherical model leads to higher surface brightness for the same reason. The wind is non-radiative even with the additional mass-loading. As we emphasize below, we have not sought a complete fit to the data in this preliminary effort. In Appendix A, we discuss how variations of wind model parameters lead to different densities, temperatures, velocities, and X-ray luminosities (see Figure \ref{fig:varied}).  

{\bf Entropy profile:} The observed entropy profile is complicated. As emphasized in Section \ref{sec:massloadedspherical}, an entropy gradient can signal mass-loading, and changes in sign of the entropy gradient imply that the hot gas Mach number is crossing the critical Mach number of $\sim1.9$ from equation (\ref{eq:Mcrit}). Observations of the temperature (sound speed), combined with a positive/negative entropy gradient can be used to put an lower/upper limit on the hot gas velocity, respectively (eq.~\ref{eq:v}). The lower left panel of Figure \ref{fig:plots} shows the measurements entropy from the data. Using $T(s)$, we see that the positive entropy gradient outside the core to the south from $s\simeq-0.5$ to $-2$\,kpc puts a lower limit on the velocity of $\sim800$\,km s$^{-1}$ that closely tracks the velocity profile obtained by the mass-loaded model (orange; upper right panel). Points of local extrema in the entropy profile where $dK/ds=0$ in principle provide measurements of the velocity, given $T$. For example at $s\simeq-2$\,kpc, $\log_{10}[T]\simeq6.9$ gives $v\simeq800$\,km s$^{-1}$. Similar statements can be made at the other inflections between positive and negative entropy gradient across the northern and southern profiles. All indicate velocities less than $\simeq900$\, km s$^{-1}$. It is notable that without appeal to the models, an empirical constraint can be set on $v$ that yields a velocity significantly below the previous predictions of \cite{Strickland2009}, who find an asymptotic velocities of $\sim1400-2200$\,km s$^{-1}$.

Turning to the model entropy profiles, we note that the mass-loaded  model (orange) is not a particularly good fit to the complicated entropy profile exhibited by the data. While we are able to produce a strong decrease in $K$ right outside the starburst core to the south because we start mass-loading when the Mach number is less than 1.9, we are unable to produce the deep drop seen in the data. As discussed below, a more thorough and systematic search through the large parameter space of these models is needed.

{\bf Important caveats:} The models shown are not fits. They were not constructed by a systematic search through parameter space, but rather by-eye and with a limited by-hand search through some of the relevant parameters. While $A(s)$ was defined by the observations, the total number of parameters for the models is $\alpha$, $\beta$, and $R$ for the core and $\dot{\mu}_0$, $R_{\rm load}$, $a$, and $\Delta=\Gamma$. A future work could embed the general expressions in a systematic search through these parameters, and with a direct comparison with the data and uncertainties with an MCMC approach. A limitation of the current numerical models is that we are solving ODEs that, with sufficient mass-loading, can encounter a sonic point, where the flow goes from supersonic to subsonic. We are not yet able to effectively deal sub/super-sonic transitions in our models and this will be the subject of a future development.

In addition, there are several physical effects we have not included in these models. Most important is the back-reaction of the hot flow on the cool clouds. The equations in Section \ref{sec:two} assume that the cool material is injected with zero net velocity. Yet, as discussed in Section \ref{sec:cgols} in our discussion of the CGOLS IV simulation, the cool cloud medium is dynamically coupled with the hot wind. Parameterized models including the dynamics of both a cool and hot phase, calibrated with numerical simulations, may provide insights beyond the limitations of a single-fluid description. 

An additional remaining uncertainty in the comparison of our models with the data is that \cite{Lopez2020} find three different thermal components, which we have modelled here as a single component (eqs.~\ref{eq:ttot} and \ref{eq:ntot}). Precisely how this should be done remains unclear and will also be the subject of future work. A possibility that we return to in the discussion below is that the three phases are not co-spatial, that the hottest phase is at the center of a cylindrical geometry and evolves in isolation, whereas the entrained cool material is provided by the confining medium, and thus the phase we model with most fidelity has a hollow cone, frustum-like geometry. 

In summary, we note several important successes and applications of the model here. The first is that a determination of $A(s)$ directly from the X-ray data (Fig.~\ref{fig:M82areas})immediately provides a geometry that gives much flatter temperature and density profiles that are much more indicative of the data. The additional inclusion of mass-loading via the equations presented in Section \ref{sec:two} yields $T$, $n$, X-ray luminosity, and composition profiles closer to those observed, but with an entropy profile we cannot readily match. Interpreted in this model, composition changes to the hot flow directly map to the composition of the swept up medium. Further, regardless of the models, inflections in the entropy profile where $dK/ds=0$ can in principle yield direct constraints on the hot gas velocity that may be probed by upcoming missions like XRISM. Taking the temperature and entropy profiles shown in Figure \ref{fig:plots} and the places where $dK/ds=0$ at face value implies hot wind velocities of $\simeq800$\,km s$^{-1}$ that are much smaller than have previously been inferred for the hot wind on the basis of the hot diffuse X-ray emission at M82's core ($\simeq1400-2200$\,km s$^{-1}$; \citealt{Strickland2009}).

\section{Discussion \& Summary}
\label{sec:five} 

In this paper we investigate how the bulk properties of a single-component hot galactic wind are modified by both mass-loading from a cool phase and non-spherical areal divergence (eqs.~\ref{eq:continuity}-\ref{eq:energy}). Because cool clouds are expected to be swept up and incorporated into the hot flow of a supersonic galactic wind, and because X-ray observations suggest that such flows do not expand spherically, this work is directly related to the interpretation of observed systems like M82, NGC 253, and other local starburst galaxies with resolved X-ray emission \citep{Strickland2000}.

In the limit that the hot flow is highly supersonic, Equations (\ref{eq:dvds2}) - (\ref{eq:dKds2}) show how both mass-loading and non-spherical divergence affect both the thermodynamics and dynamics: mass-loading generically decelerates the hot flow, increases the density and temperature, flattening these gradients, and leads to a positive entropy gradient if the Mach number is large. This combination is a hallmark of mass-loading in a hot supersonic wind. The flow geometry can lead to deceleration/acceleration or increases/decreases in density and pressure, depending on the shape of the areal function $A(s)$, but does not affect the entropy. For a specified flow geometry (e.g., as determined by observations), the volumetric mass-loading rate can be directly inferred by applying the general relations. 

We present examples for flows undergoing prescribed expansion geometries in Figures \ref{fig:flaredcyl} and \ref{fig:flutedcyl}. 
For a flow geometry that undergoes rapid changes, as in the fluted-cylinder case study (Section \ref{sec:nonsphericaladiabatic}), the criterion given by Equation \ref{eq:decelGeo} may be met such that deceleration can occur despite there being no mass-loading. Equation \ref{eq:decelGeo} provides a criterion for significant acceleration (and deceleration) of a supersonic flow solely due to a rapidly changing flow geometry. 

Regardless of the flow geometry, the entropy profile is indicative of whether mass-loading occurs or not. Mass-loading partially thermalizes the kinetic energy of the hot flow, which provides heating and changes the entropy profile (Equation \ref{eq:dlnKds}). When a supersonic flow is mass-loaded, the entropy increases if the Mach number is larger than a critical value $\mathcal{M} \geq \mathcal{M}_\mathrm{K,crit}$ and decreases for $\mathcal{M} < \mathcal{M}_\mathrm{K,crit}$, where $\mathcal{M}_\mathrm{K,crit}\simeq1.9$ for $\gamma=5/3$ is given by equation (\ref{eq:Mcrit}). In the limit that radiative cooling does not significantly affect the entropy profile, observed entropy gradients can be used to set a maximum or minimum on the hot gas velocity (see Equation \ref{eq:v}). As we note in our discussion of M82 in Section \ref{sec:m82} and Figure \ref{fig:plots}, in principle, extrema in the observed entropy profile $dK/ds=0$ provide a measurement of the hot gas Mach number (see below).

In Section \ref{sec:cgols}, we apply our results on mass loading to the high-resolution CGOLS IV simulation \citep{Schneider2020}. At 35 Myr, the hot gas temperature, density, and pressure radial profiles do not follow the expectations for spherical adiabatic expansion. While \cite{Schneider2020}  fit the density profile by hand with a power-law, they did not match the temperature, pressure, and entropy profiles. In Section \ref{sec:cgols} we consider whether mixing from the cool phase into the hot phase explains the profiles from the CGOLS IV simulation. Using Equation \ref{eq:continuity}, we infer the rate of cold gas entrainment (Fig.~\ref{fig:infermudot}) and then use these profiles to solve the coupled wind equations for mass-loaded models. In Figure \ref{fig:cgols}, we show the the models simultaneously capture the temperature, density, and pressure profiles for $r \gtrsim 1$\,kpc. Furthermore, we match additional quantities (see Figure \ref{fig:cgols}) such as the entropy profile and decreasing scalar variable $c$, which tracks how much mass is mixed into the hot flow. In employing a simple CC85 wind model as the initial condition of the problem, we do not capture the complicated cluster feedback mechanism used in CGOLS IV near the base of the outflow. Overall, our models reproduce the flat temperature, density, entropy, and pressure profiles seen from the CGOLS IV simulation well, providing strong evidence that mixing from the cool phase directly shapes the bulk hot gas wind.


We also apply our results to the recent analysis of M82 by \citet{Lopez2020}, which revealed hot gas temperature and density profiles that fall off much more slowly than the expectation for spherical adiabatic expansion. We suggest that these profiles indicate both non-spherical expansion and mixing of cold material into the hot phase. The area function that describes the non-spherical geometry is directly extracted from transverse surface brightness profiles (see Section \ref{sec:m82} and Fig.~\ref{fig:M82areas}). With this area function, we show that a model without mass loading (blue line in Figure \ref{fig:plots}) does a much better job at matching the temperature and density profiles than a spherical model without mass-loading (dash-dotted gray line). By including mass-loading (orange line), the temperatures, densities, and luminosities are better matched. We find that the scalar variable that tracks the amount of mixing qualitatively matched the Si abundance profile (black data points), but not the Fe or Si abundances (light gray), perhaps indicating the relative abundance of the swept up gas in these elements with respect to that injected in the core (see Section \ref{sec:m82}). The asymptotic velocities of the mass-loaded non-spherical wind model (orange line) are $\sim$900\,km s$^{-1}$ and $\sim$1300\,km s$^{-1}$ to the south and north, respectively. This is substantially lower than the $\sim$1600\,km s$^{-1}$ velocities predicted by a spherical adiabatic model \citep{Strickland2009}. As discussed in Sections \ref{sec:entropygrad} and \ref{sec:m82}, the entropy gradient of a supersonic mass-loaded wind provides direct constraints on the bulk hot gas velocity. Figures \ref{fig:M82areas} and \ref{fig:plots} show that the area function extracted from the transverse surface brightness profiles heavily influences the evolution of the flow. A higher spatial resolution study of M82 is required for a more accurate characterization of the flow tube $A(s)$. Higher resolution will also better pinpoint where along the flow $dK/ds=0$, which provides a measurement of hot gas velocity (Equation \ref{eq:v}). These topics will be studied in a future work for M82.

The X-ray Imaging and Spectroscopy Mission (XRISM) is an upcoming mission with 20-40 times better spectral resolution than current CCD instruments used on Chandra, XMM-Newton, and Suzaku \citep{XRISM2020}. This, and the substantially increased collecting area and bandpass over the grating instruments on those missions mean that XRISM is capable and expected to make the first measurements of hot X-ray emitting gas kinematics with an accuracty of $<100$\,km\,s$^{-1}$. The mass-loaded wind models presented in this work make a testable prediction for the forthcoming hot gas velocity measurements by XRISM.

In making more detailed comparisons with the X-ray surface brightness profiles of observed systems, we note that these often exhibit limb brightening at the ``edges" of the hot flow. This is evident by eye in the X-ray surface brightness profile of the ``frustum" in \cite{Strickland2000}'s discussion of the starburst NGC 253 and in NGC 3079 \citep{Hodges-Kluck2020}, among others \citep{Strickland2004a,Strickland2004b}.  In Figure \ref{fig:plots}, we show that mass-loading  increases the density and thus the emissivity of the X-ray emitting medium. Thus, the mass-loading phenomenon provides a natural explanation for the limb-brightening seen in the X-ray profiles if the mixing is occurring at the ``edge" of $A(s)$ in a frustum-like geometry. Here, the picture would be that a hot un-mass-loaded wind emerges from the inner region, directly along the minor axis, while the mass-loading and mixing occurs in a brighter cocoon or sheath provided by the cooler mixing material, as sketched in Figure 19 by \cite{Leroy2015}. 

There are a number of directions for future theoretical work. First, while we have focused on the super-sonic region of hot galactic winds, strong enough mass-loading can force the flow to go through a sonic point (making the hot flow transition from super-sonic to sub-sonic), which changes the character of the flow solutions. Second, strong deceleration may also cause time-dependent instabilities that our steady-state solutions do not capture \citep{Schekinov1996}. Third, mass-loaded hot subsonic and supersonic flows should be investigated to see if they provide thermodynamic backgrounds more conducive to cool cloud survival and growth (e.g., \citealt{Gronke2018}). Fourth, a critical ingredient for future investigations on the bulk properties of mass-loaded hot winds (see Sections \ref{sec:cgols} and \ref{sec:m82}) would be to consistently include a model of the cool medium, perhaps in the form of a spectrum of clouds \citep{Suchkov1996}, so that the back-reaction of the hot medium on the cool clouds can be calculated. In our models so far, we have approximated the cool swept-up medium as having zero net velocity, but a model incorporating momentum equations for a cool medium could be used to capture these dynamics \citep{Suchkov1996,Fielding2021}.

\section*{Data Availability}
The data underlying this article will be shared on request to the corresponding author.

\section*{Acknowledgements}
TAT and DDN thank Evan Schneider and Laura Lopez for providing CGOLS IV and M82 data, respectively, and the Ohio State Astronomy Gas, Galaxies, and Feedback Group for illuminating conversations. TAT thanks Eliot Quataert, Eve Ostriker, Adam Leroy, Evan Schneider, and Brant Roberson for discussions. TAT acknowledges support from a Simons Foundation Fellowship in Theoretical Physics and an IBM Einstein Fellowship from the Institute for Advanced Study, Princeton, during the time when part of this work was completed. DDN and TAT are supported by National Science Foundation Grant \#1516967 and NASA ATP 80NSSC18K0526. 

\bibliography{bibliography}

\begin{thebibliography}{}

\bibitem[\protect\citeauthoryear{{Aguirre}, {Hernquist}, {Schaye}, {Katz},
  {Weinberg} \& {Gardner}}{{Aguirre} et~al.}{2001}]{Aguirre2001}
{Aguirre} A.,  {Hernquist} L.,  {Schaye} J.,  {Katz} N.,  {Weinberg} D.~H.,
  {Gardner} J.,  2001, \apj, 561, 521

\bibitem[\protect\citeauthoryear{{Anders} \& {Grevesse}}{{Anders} \&
  {Grevesse}}{1989}]{Anders1989}
{Anders} E.,  {Grevesse} N.,  1989, gca, 53, 197

\bibitem[\protect\citeauthoryear{{Breitschwerdt}, {McKenzie} \&
  {Voelk}}{{Breitschwerdt} et~al.}{1991}]{Breitschwerdt1991}
{Breitschwerdt} D.,  {McKenzie} J.~F.,    {Voelk} H.~J.,  1991, \aap, 245, 79

\bibitem[\protect\citeauthoryear{{Chevalier} \& {Clegg}}{{Chevalier} \&
  {Clegg}}{1985}]{Chevalier1985}
{Chevalier} R.~A.,  {Clegg} A.~W.,  1985, \nat, 317, 44

\bibitem[\protect\citeauthoryear{{Cooper}, {Bicknell}, {Sutherland } \&
  {Bland-Hawthorn}}{{Cooper} et~al.}{2009}]{Cooper2009}
{Cooper} J.~L.,  {Bicknell} G.~V.,  {Sutherland } R.~S.,    {Bland-Hawthorn}
  J.,  2009, \apj, 703, 330

\bibitem[\protect\citeauthoryear{{Cowie}, {McKee} \& {Ostriker}}{{Cowie}
  et~al.}{1981}]{Cowie1981}
{Cowie} L.~L.,  {McKee} C.~F.,    {Ostriker} J.~P.,  1981, \apj, 247, 908

\bibitem[\protect\citeauthoryear{{Dekel} \& {Silk}}{{Dekel} \&
  {Silk}}{1986}]{Dekel1986}
{Dekel} A.,  {Silk} J.,  1986, \apj, 303, 39

\bibitem[\protect\citeauthoryear{{Everett}, {Zweibel}, {Benjamin}, {McCammon},
  {Rocks} \& {Gallagher} III}{{Everett} et~al.}{2008}]{Everett2008}
{Everett} J.~E.,  {Zweibel} E.~G.,  {Benjamin} R.~A.,  {McCammon} D.,  {Rocks}
  L.,    {Gallagher} III J.~S.,  2008, \apj, 674, 258

\bibitem[\protect\citeauthoryear{{Ferland}, {Porter}, {van Hoof}, {Williams},
  {Abel}, {Lykins}, {Shaw}, {Henney} \& {Stancil}}{{Ferland}
  et~al.}{2013}]{Ferland2013}
{Ferland} G.~J.,  {Porter} R.~L.,  {van Hoof} P.~A.~M.,  {Williams} R.~J.~R.,
  {Abel} N.~P.,  {Lykins} M.~L.,  {Shaw} G.,  {Henney} W.~J.,    {Stancil}
  P.~C.,  2013, rmxaa, 49, 137

\bibitem[\protect\citeauthoryear{{Fielding} \& {Bryan}}{{Fielding} \&
  {Bryan}}{2021}]{Fielding2021}
{Fielding} D.~B.,  {Bryan} G.~L.,  2021, arXiv e-prints, p. arXiv:2108.05355

\bibitem[\protect\citeauthoryear{{Finlator} \& {Dav{\'e}}}{{Finlator} \&
  {Dav{\'e}}}{2008}]{Finlator2008}
{Finlator} K.,  {Dav{\'e}} R.,  2008, \mnras, 385, 2181

\bibitem[\protect\citeauthoryear{{Foster} \& {Heuer}}{{Foster} \&
  {Heuer}}{2020}]{Foster2020}
{Foster} A.~R.,  {Heuer} K.,  2020, Atoms, 8, 49

\bibitem[\protect\citeauthoryear{{Gray}, {Oey}, {Silich} \&
  {Scannapieco}}{{Gray} et~al.}{2019}]{Gray2019b}
{Gray} W.~J.,  {Oey} M.~S.,  {Silich} S.,    {Scannapieco} E.,  2019, \apj,
  887, 161

\bibitem[\protect\citeauthoryear{{Gray}, {Scannapieco} \& {Lehnert}}{{Gray}
  et~al.}{2019}]{Gray2019a}
{Gray} W.~J.,  {Scannapieco} E.,    {Lehnert} M.~D.,  2019, \apj, 875, 110

\bibitem[\protect\citeauthoryear{{Gronke} \& {Oh}}{{Gronke} \&
  {Oh}}{2018}]{Gronke2018}
{Gronke} M.,  {Oh} S.~P.,  2018, \mnras, 480, L111

\bibitem[\protect\citeauthoryear{{Heckman}, {Alexandroff}, {Borthakur},
  {Overzier} \& {Leitherer}}{{Heckman} et~al.}{2015}]{Heckman2015}
{Heckman} T.~M.,  {Alexandroff} R.~M.,  {Borthakur} S.,  {Overzier} R.,
  {Leitherer} C.,  2015, \apj, 809, 147

\bibitem[\protect\citeauthoryear{{Heckman}, {Armus} \& {Miley}}{{Heckman}
  et~al.}{1990}]{Heckman1990}
{Heckman} T.~M.,  {Armus} L.,    {Miley} G.~K.,  1990, \apjs, 74, 833

\bibitem[\protect\citeauthoryear{{Heckman} \& {Borthakur}}{{Heckman} \&
  {Borthakur}}{2016}]{Heckman2016}
{Heckman} T.~M.,  {Borthakur} S.,  2016, \apj, 822, 9

\bibitem[\protect\citeauthoryear{{Hodges-Kluck}, {Yukita}, {Tanner}, {Ptak},
  {Bregman} \& {Li}}{{Hodges-Kluck} et~al.}{2020}]{Hodges-Kluck2020}
{Hodges-Kluck} E.~J.,  {Yukita} M.,  {Tanner} R.,  {Ptak} A.~F.,  {Bregman}
  J.~N.,    {Li} J.-t.,  2020, \apj, 903, 35

\bibitem[\protect\citeauthoryear{{Klein}, {McKee} \& {Colella}}{{Klein}
  et~al.}{1994}]{Klein1994}
{Klein} R.~I.,  {McKee} C.~F.,    {Colella} P.,  1994, \apj, 420, 213

\bibitem[\protect\citeauthoryear{{Kopp} \& {Holzer}}{{Kopp} \&
  {Holzer}}{1976}]{Kopp1976}
{Kopp} R.~A.,  {Holzer} T.~E.,  1976, solphys, 49, 43

\bibitem[\protect\citeauthoryear{{Leroy}, {Walter}, {Martini}, {Roussel},
  {Sandstrom}, {Ott}, {Weiss}, {Bolatto}, {Schuster} \&
  {Dessauges-Zavadsky}}{{Leroy} et~al.}{2015}]{Leroy2015}
{Leroy} A.~K.,  {Walter} F.,  {Martini} P.,  {Roussel} H.,  {Sandstrom} K.,
  {Ott} J.,  {Weiss} A.,  {Bolatto} A.~D.,  {Schuster} K.,
  {Dessauges-Zavadsky} M.,  2015, \apj, 814, 83

\bibitem[\protect\citeauthoryear{{Lochhaas}, {Thompson} \&
  {Schneider}}{{Lochhaas} et~al.}{2020}]{Lochhaas2020}
{Lochhaas} C.,  {Thompson} T.~A.,    {Schneider} E.~E.,  2020, arXiv e-prints,
  p. arXiv:2011.06004

\bibitem[\protect\citeauthoryear{{Lopez}, {Mathur}, {Nguyen}, {Thompson} \&
  {Olivier}}{{Lopez} et~al.}{2020}]{Lopez2020}
{Lopez} L.~A.,  {Mathur} S.,  {Nguyen} D.~D.,  {Thompson} T.~A.,    {Olivier}
  G.~M.,  2020, \apj, 904, 152

\bibitem[\protect\citeauthoryear{{Martin}}{{Martin}}{1998}]{Martin1998}
{Martin} C.~L.,  1998, \apj, 506, 222

\bibitem[\protect\citeauthoryear{{Martin}}{{Martin}}{2005}]{Martin2005}
{Martin} C.~L.,  2005, \apj, 621, 227

\bibitem[\protect\citeauthoryear{{Oppenheimer} \& {Dav{\'e}}}{{Oppenheimer} \&
  {Dav{\'e}}}{2006}]{Oppenheimer2006}
{Oppenheimer} B.~D.,  {Dav{\'e}} R.,  2006, \mnras, 373, 1265

\bibitem[\protect\citeauthoryear{{Oppenheimer} \& {Dav{\'e}}}{{Oppenheimer} \&
  {Dav{\'e}}}{2008}]{Oppenheimer2008}
{Oppenheimer} B.~D.,  {Dav{\'e}} R.,  2008, \mnras, 387, 577

\bibitem[\protect\citeauthoryear{{Peeples} \& {Shankar}}{{Peeples} \&
  {Shankar}}{2011}]{Peeples2011}
{Peeples} M.~S.,  {Shankar} F.,  2011, \mnras, 417, 2962

\bibitem[\protect\citeauthoryear{{Pettini}, {Shapley}, {Steidel}, {Cuby},
  {Dickinson}, {Moorwood}, {Adelberger} \& {Giavalisco}}{{Pettini}
  et~al.}{2001}]{Pettini2001}
{Pettini} M.,  {Shapley} A.~E.,  {Steidel} C.~C.,  {Cuby} J.-G.,  {Dickinson}
  M.,  {Moorwood} A. F.~M.,  {Adelberger} K.~L.,    {Giavalisco} M.,  2001,
  \apj, 554, 981

\bibitem[\protect\citeauthoryear{{Ranalli}, {Comastri}, {Origlia} \&
  {Maiolino}}{{Ranalli} et~al.}{2008}]{Ranalli2008}
{Ranalli} P.,  {Comastri} A.,  {Origlia} L.,    {Maiolino} R.,  2008, \mnras,
  386, 1464

\bibitem[\protect\citeauthoryear{{Rubin}, {Weiner}, {Koo}, {Martin},
  {Prochaska}, {Coil} \& {Newman}}{{Rubin} et~al.}{2010}]{Rubin2010}
{Rubin} K. H.~R.,  {Weiner} B.~J.,  {Koo} D.~C.,  {Martin} C.~L.,  {Prochaska}
  J.~X.,  {Coil} A.~L.,    {Newman} J.~A.,  2010, \apj, 719, 1503

\bibitem[\protect\citeauthoryear{{Scannapieco} \& {Br{\"u}ggen}}{{Scannapieco}
  \& {Br{\"u}ggen}}{2015}]{Scannapieco2015}
{Scannapieco} E.,  {Br{\"u}ggen} M.,  2015, \apj, 805, 158

\bibitem[\protect\citeauthoryear{{Scannapieco}, {Ferrara} \&
  {Madau}}{{Scannapieco} et~al.}{2002}]{Scannapieco2002}
{Scannapieco} E.,  {Ferrara} A.,    {Madau} P.,  2002, \apj, 574, 590

\bibitem[\protect\citeauthoryear{{Schneider}, {Ostriker}, {Robertson} \&
  {Thompson}}{{Schneider} et~al.}{2020}]{Schneider2020}
{Schneider} E.~E.,  {Ostriker} E.~C.,  {Robertson} B.~E.,    {Thompson} T.~A.,
  2020, \apj, 895, 43

\bibitem[\protect\citeauthoryear{{Schneider}, {Robertson} \&
  {Thompson}}{{Schneider} et~al.}{2018}]{Schneider2018}
{Schneider} E.~E.,  {Robertson} B.~E.,    {Thompson} T.~A.,  2018, \apj, 862,
  56

\bibitem[\protect\citeauthoryear{{Shchekinov}}{{Shchekinov}}{1996}]{Schekinov1996}
{Shchekinov} Y.~A.,  1996, Geophysical and Astrophysical Fluid Dynamics, 82, 69

\bibitem[\protect\citeauthoryear{{Silich}, {Tenorio-Tagle} \&
  {Mu{\~n}oz-Tu{\~n}{\'o}n}}{{Silich} et~al.}{2003}]{Silich2003}
{Silich} S.,  {Tenorio-Tagle} G.,    {Mu{\~n}oz-Tu{\~n}{\'o}n} C.,  2003, \apj,
  590, 791

\bibitem[\protect\citeauthoryear{{Silich}, {Tenorio-Tagle} \&
  {Rodr{\'\i}guez-Gonz{\'a}lez}}{{Silich} et~al.}{2004}]{Silich2004}
{Silich} S.,  {Tenorio-Tagle} G.,    {Rodr{\'\i}guez-Gonz{\'a}lez} A.,  2004,
  \apj, 610, 226

\bibitem[\protect\citeauthoryear{{Strickland} \& {Heckman}}{{Strickland} \&
  {Heckman}}{2007}]{Strickland2007}
{Strickland} D.~K.,  {Heckman} T.~M.,  2007, \apj, 658, 258

\bibitem[\protect\citeauthoryear{{Strickland} \& {Heckman}}{{Strickland} \&
  {Heckman}}{2009}]{Strickland2009}
{Strickland} D.~K.,  {Heckman} T.~M.,  2009, \apj, 697, 2030

\bibitem[\protect\citeauthoryear{{Strickland}, {Heckman}, {Colbert}, {Hoopes}
  \& {Weaver}}{{Strickland} et~al.}{2004b}]{Strickland2004b}
{Strickland} D.~K.,  {Heckman} T.~M.,  {Colbert} E. J.~M.,  {Hoopes} C.~G.,
  {Weaver} K.~A.,  2004b, \apjs, 151, 193

\bibitem[\protect\citeauthoryear{{Strickland}, {Heckman}, {Colbert}, {Hoopes}
  \& {Weaver}}{{Strickland} et~al.}{2004a}]{Strickland2004a}
{Strickland} D.~K.,  {Heckman} T.~M.,  {Colbert} E. J.~M.,  {Hoopes} C.~G.,
  {Weaver} K.~A.,  2004a, \apj, 606, 829

\bibitem[\protect\citeauthoryear{{Strickland}, {Heckman}, {Weaver} \&
  {Dahlem}}{{Strickland} et~al.}{2000}]{Strickland2000}
{Strickland} D.~K.,  {Heckman} T.~M.,  {Weaver} K.~A.,    {Dahlem} M.,  2000,
  \aj, 120, 2965

\bibitem[\protect\citeauthoryear{{Suchkov}, {Berman}, {Heckman} \&
  {Balsara}}{{Suchkov} et~al.}{1996}]{Suchkov1996}
{Suchkov} A.~A.,  {Berman} V.~G.,  {Heckman} T.~M.,    {Balsara} D.~S.,  1996,
  \apj, 463, 528

\bibitem[\protect\citeauthoryear{{Thompson}, {Quataert}, {Zhang} \&
  {Weinberg}}{{Thompson} et~al.}{2016}]{Thompson2016}
{Thompson} T.~A.,  {Quataert} E.,  {Zhang} D.,    {Weinberg} D.~H.,  2016,
  \mnras, 455, 1830

\bibitem[\protect\citeauthoryear{{Tremonti}, {Heckman}, {Kauffmann},
  {Brinchmann}, {Charlot}, {White}, {Seibert}, {Peng}, {Schlegel}, {Uomoto},
  {Fukugita} \& {Brinkmann}}{{Tremonti} et~al.}{2004}]{Tremonti2004}
{Tremonti} C.~A.,  {Heckman} T.~M.,  {Kauffmann} G.,  {Brinchmann} J.,
  {Charlot} S.,  {White} S.~D.~M.,  {Seibert} M.,  {Peng} E.~W.,  {Schlegel}
  D.~J.,  {Uomoto} A.,  {Fukugita} M.,    {Brinkmann} J.,  2004, \apj, 613, 898

\bibitem[\protect\citeauthoryear{{Walter}, {Weiss} \& {Scoville}}{{Walter}
  et~al.}{2002}]{Walter2002}
{Walter} F.,  {Weiss} A.,    {Scoville} N.,  2002, \apjl, 580, L21

\bibitem[\protect\citeauthoryear{{Wang}}{{Wang}}{1995}]{Wang1995}
{Wang} B.,  1995, \apj, 444, 590

\bibitem[\protect\citeauthoryear{{Westmoquette}, {Gallagher}, {Smith},
  {Trancho}, {Bastian} \& {Konstantopoulos}}{{Westmoquette}
  et~al.}{2009}]{Westmoquette2009}
{Westmoquette} M.~S.,  {Gallagher} J.~S.,  {Smith} L.~J.,  {Trancho} G.,
  {Bastian} N.,    {Konstantopoulos} I.~S.,  2009, \apj, 706, 1571

\bibitem[\protect\citeauthoryear{{W{\"u}nsch}, {Silich}, {Palou{\v{s}}} \&
  {Tenorio-Tagle}}{{W{\"u}nsch} et~al.}{2007}]{Wunsch2007}
{W{\"u}nsch} R.,  {Silich} S.,  {Palou{\v{s}}} J.,    {Tenorio-Tagle} G.,
  2007, \aap, 471, 579

\bibitem[\protect\citeauthoryear{{W{\"u}nsch}, {Tenorio-Tagle}, {Palou{\v{s}}}
  \& {Silich}}{{W{\"u}nsch} et~al.}{2008}]{Wunsch2008}
{W{\"u}nsch} R.,  {Tenorio-Tagle} G.,  {Palou{\v{s}}} J.,    {Silich} S.,
  2008, \apj, 683, 683

\bibitem[\protect\citeauthoryear{{XRISM Science Team}}{{XRISM Science
  Team}}{2020}]{XRISM2020}
{XRISM Science Team} 2020, arXiv e-prints, p. arXiv:2003.04962

\bibitem[\protect\citeauthoryear{{Yu}, {Owen}, {Wu} \& {Ferreras}}{{Yu}
  et~al.}{2020}]{Yu2020}
{Yu} B.~P.~B.,  {Owen} E.~R.,  {Wu} K.,    {Ferreras} I.,  2020, \mnras, 492,
  3179

\bibitem[\protect\citeauthoryear{{Zhang}, {Thompson}, {Murray} \&
  {Quataert}}{{Zhang} et~al.}{2014}]{Zhang2014}
{Zhang} D.,  {Thompson} T.~A.,  {Murray} N.,    {Quataert} E.,  2014, \apj,
  784, 93

\bibitem[\protect\citeauthoryear{{Zhang}, {Thompson}, {Quataert} \&
  {Murray}}{{Zhang} et~al.}{2017}]{Zhang2017}
{Zhang} D.,  {Thompson} T.~A.,  {Quataert} E.,    {Murray} N.,  2017, \mnras,
  468, 4801

\end{thebibliography}

\section*{Appendix A: Varied Wind Model Parameters for M82}
\begin{figure*}
    \centering
    \includegraphics[width=\textwidth]{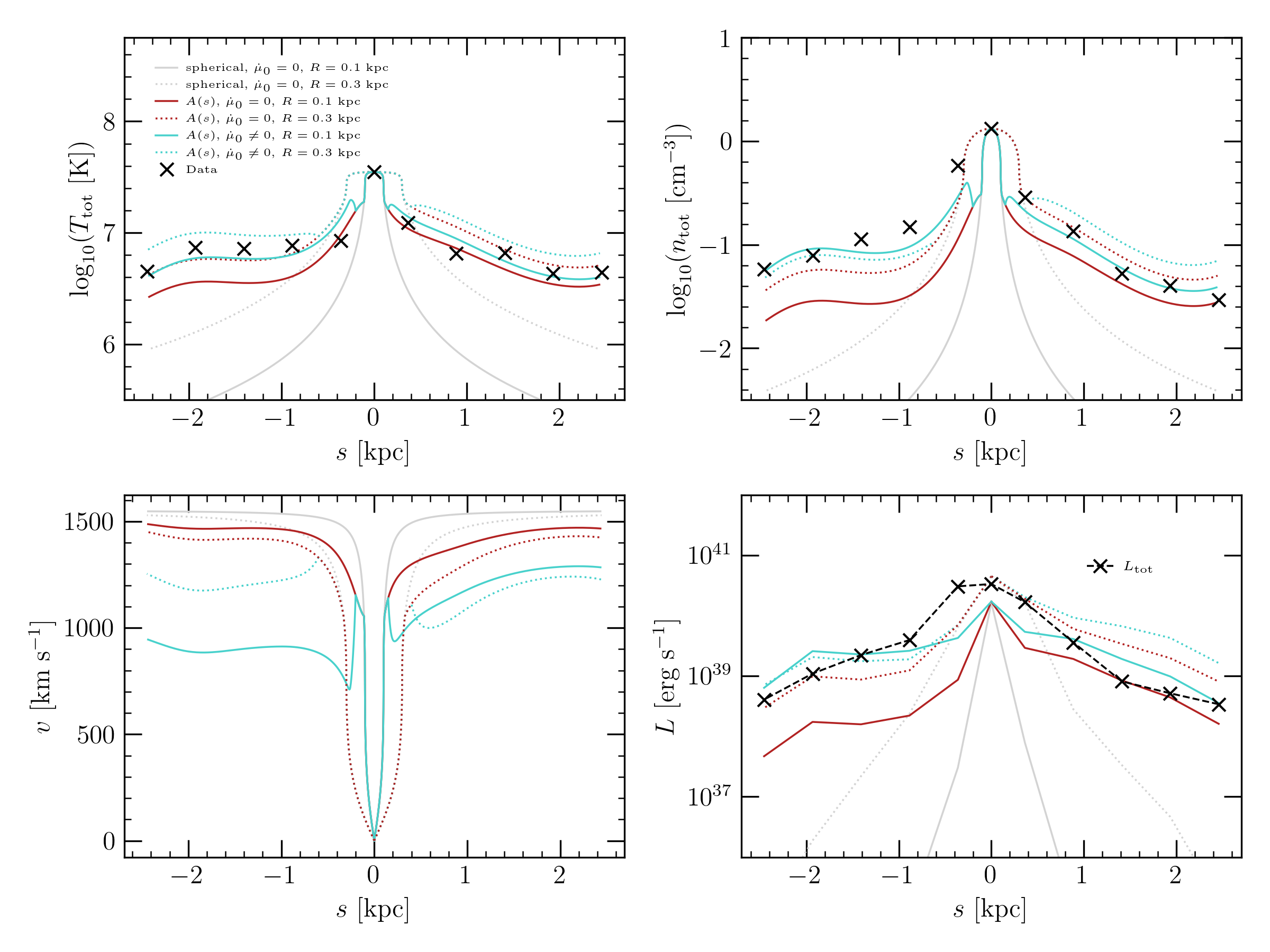}
    \caption{Temperature, density, velocity, and luminosity profiles for six different wind models: $R=0.1$\,kpc (solid lines), $R=0.3$\,kpc (dotted lines), spherical (gray lines), non-spherical without mass-loading (red lines), and non-spherical with mass-loading (turquoise lines). The black x's are the total temperature and density of the three temperature components observed by Chandra (private communication, L. Lopez).} 
    \label{fig:varied}
\end{figure*}

For fixed $\dot{M}_\mathrm{SFR}$, after fitting $\alpha$ and $\beta$ to the central temperature and density (Equations \ref{eq:Tc} and \ref{eq:nc}) and using the $A(s)$ derived from the transverse X-ray surface brightness profiles (see Figure \ref{fig:M82areas}), there are 6 parameters (for total of 12, for both the north and south wind axes) to consider: $R$, $\dot{\mu}_0$, $\Delta$, $\Gamma$, $a$, and $R_\mathrm{load}$ (see Equation \ref{eq:mudot2power}). In Figure \ref{fig:varied} we consider several choices for wind model parameters for illustrative purposes, which can be contrasted with Figure \ref{fig:plots}. Solid and dotted lines have a starburst radius of $R=0.1$ and $R=0.3$\,kpc, respectively. The gray lines are spherical models without mass-loading, red lines are non-spherical models without mass-loading, and the cyan lines are non-spherical wind models with mass-loading. The solid cyan line has mass-loading parameters of $R_\mathrm{load}=0.2$\, kpc, $\dot{\mu}_0=0.36$\,$\mathrm{M_\odot \ yr^{-1}}$, $\Delta=\Gamma=3.5$, and $a=0.7$\,kpc and $R_\mathrm{load}=0.15$\, kpc, $\dot{\mu}_0=0.01$\,$\mathrm{M_\odot \ yr^{-1}}$, $\Delta=\Gamma=5$, and $a=0.7$\,kpc for the southern and northern sides, respectively. The dotted cyan line has mass-loading parameters of $R_\mathrm{load}=0.6$\, kpc, $\dot{\mu}_0=2.0$\,$\mathrm{M_\odot \ yr^{-1}}$, $\Delta=3.5$, $\Gamma=5.0$, and $a=0.7$\,kpc and $R_\mathrm{load}=0.4$\, kpc, $\dot{\mu}_0=0.1$\,$\mathrm{M_\odot \ yr^{-1}}$, $\Delta=5$, $\Gamma=2.5$, and $a=0.7$\,kpc for the southern and northern sides, respectively. 

The entropy decreases due to mass loading for the solid cyan line (southern), whereas the entropy increases for the dotted cyan line. In the solid line wind model, the gas is decelerated such that the Mach number drops below $\mathcal{M}_\mathrm{crit}$ (Equation \ref{eq:Mcrit}) so that the entropy decreases due to mass-loading. In the dotted cyan line wind model, while the gas is decelerated due to mass loaded, this criterion is not met and the entropy increases. In Figure \ref{fig:varied}, the cyan lines have different power-law injection slopes which leads to different velocity profiles. As we would expect, models with steeper mass-loading rates (parameterized by $\Delta$, see eq.~\ref{eq:mudot2power}) lead to sharper deceleration, as more mass is being deposited over a smaller volume. Where mass-loading occurs is also important. If mass-loading occurs right outside of the starburst boundary, the flow is close to Mach 1, and a sonic point is quickly encountered. As we have discussed throughout the paper, more mass-loading lead to increased temperature, density, and luminosity, and slower velocity. Flatter-than-spherical geometries, as inferred from the M82 data, lead to flatter temperature and density gradients, and higher surface brightness, but similar velocities (bottom left, red lines, Figure \ref{fig:varied}).

\section*{Appendix B: Cooling Function}
\begin{figure}
    \centering
    \includegraphics[width=\columnwidth]{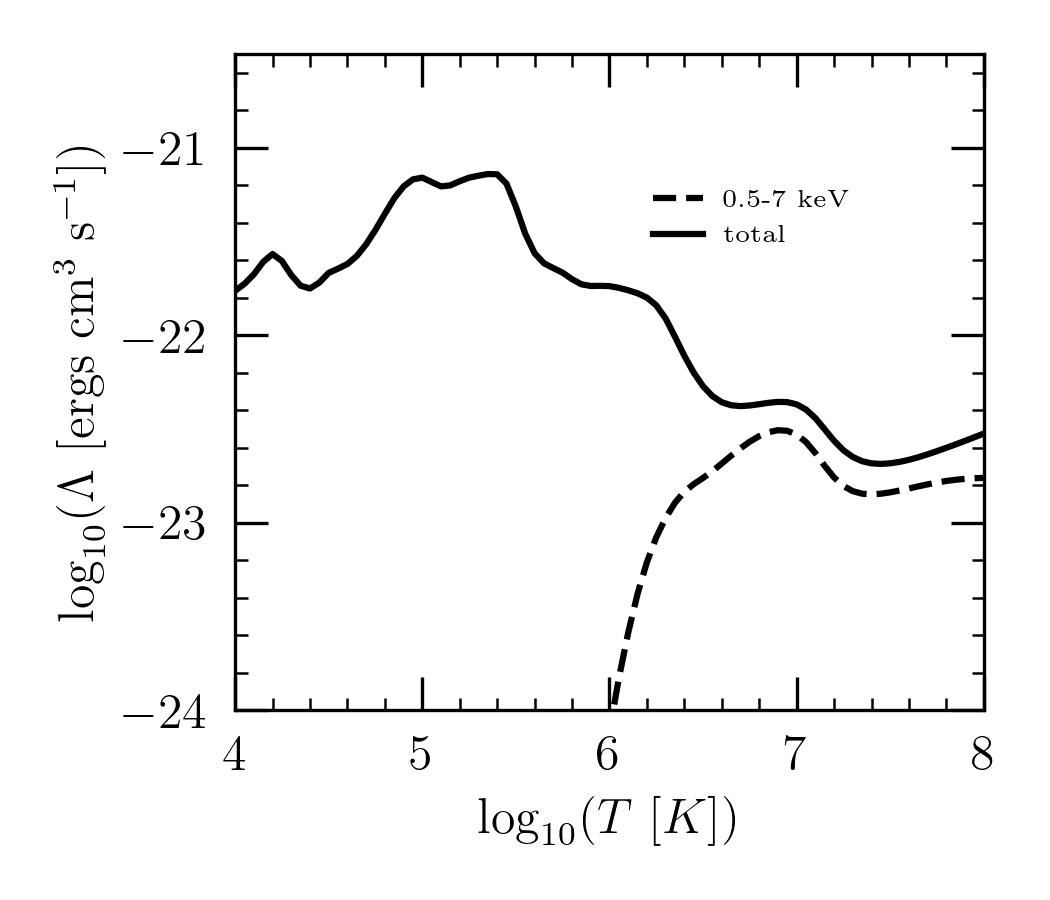}
    \caption{Cooling functions for a solar metallicity plasma in CIE for two different energy bands calculated with \texttt{PyAtomDB} \citep{Foster2020}. The total band is calculated by integrating the emissivity from $10^{-2}$ to $10^2$ keV. The 0.5 to 7 keV bandwidth corresponds to the observing range of \citet{Lopez2020}.}
    \label{fig:ct}
\end{figure}

In Figure \ref{fig:ct} we plot the total and the observing band (0.5-7\,keV) cooling functions we use to calculate the luminosities in Figures \ref{fig:plots} and \ref{fig:varied}. We use the \texttt{PyAtomDB} package to determine $\Lambda(T,\nu)$ of a hot plasma in CIE. \texttt{PyAtomDB} uses built-in tables from \citet{Anders1989} for a solar metallicity plasma. We note the breakdown of ionization equilibrium may significantly change the expected luminosity because the medium can become over-ionized relative to the local temperature for fast enough expansion. \cite{Gray2019a} find that over a wide range of parameters, far from the starburst core ($r\gg R$), the wind models generally have distributions in their ionization states that are skewed to higher ionization states compared to models that assume photoionization equilibrium. \cite{Gray2019b} show that winds that undergo rapid cooling produce strong nebular line emission compared to adiabatic outflows. They show that such models exhibit non-equilibrium conditions, thereby generating more highly ionized states than equivalent equilibrium models. 

\end{document}